%% file: SOVC_Reddit.tex
\documentclass[acmsmall]{acmart}

\AtBeginDocument{%
  \providecommand\BibTeX{{%
    \normalfont B\kern-0.5em{\scshape i\kern-0.25em b}\kern-0.8em\TeX}}}

\setcopyright{acmcopyright}
\copyrightyear{2018}
\acmYear{2018}
\acmDOI{XXXXXXX.XXXXXXX}

\acmConference[Conference acronym 'XX]{Make sure to enter the correct
  conference title from your rights confirmation emai}{June 03--05,
  2018}{Woodstock, NY}
\acmPrice{15.00}
\acmISBN{978-1-4503-XXXX-X/18/06}

\usepackage{graphicx}
\usepackage{hyperref}

\usepackage{multirow}
\usepackage{booktabs}
\usepackage{subcaption}
\usepackage{array}
\newcolumntype{L}[1]{>{\raggedright\let\newline\\\arraybackslash\hspace{0pt}}m{#1}}
\newcolumntype{C}[1]{>{\centering\let\newline\\\arraybackslash\hspace{0pt}}m{#1}}
\newcolumntype{R}[1]{>{\raggedleft\let\newline\\\arraybackslash\hspace{0pt}}m{#1}}

\begin{document}

\title[Community Archetypes]{Community Archetypes: An Empirical Framework for Guiding Research Methodologies to Reflect User Experiences of Sense of Virtual Community}

% \author{Anonymous}
% \renewcommand{\shortauthors}{Anonymous}

\author{Gale H. Prinster}
\authornote{Both authors contributed equally to this research.}
\email{abpr9862@colorado.edu}
\orcid{0000-0002-7018-8657}
\affiliation{%
  \institution{University of Colorado Boulder}
  \city{Boulder}
  \state{Colorado}
  \country{USA}
  \postcode{80304}
}
\author{C. Estelle Smith}
\authornotemark[1]
\email{estellesmith@mines.edu}
\orcid{0000-0002-4981-7105}
\affiliation{%
  \institution{Colorado School of Mines}
  \streetaddress{}
  \city{Golden}
  \state{CO}
  \country{USA}
  \postcode{80401}
}

\author{Chenhao Tan}
\orcid{0000-0002-3981-2116}
\affiliation{%
  \institution{University of Chicago}
  \city{Chicago}
  \state{Illinois}
  \country{USA}
  \postcode{80304}
}
\email{chenhao@uchicago.edu}

\author{Brian C. Keegan}
\orcid{0000-0002-7793-398X}
\affiliation{%
  \institution{University of Colorado Boulder}
  \city{Boulder}
  \state{Colorado}
  \country{USA}
  \postcode{80304}
}
\email{brian.keegan@colorado.edu}

\renewcommand{\shortauthors}{Gale H. Prinster and C. Estelle Smith, \textit{et al.}}

\begin{abstract}
Humans need a sense of community (SOC), and social media platforms afford opportunities to address this need by providing users with a sense of \textit{virtual} community (SOVC). This paper explores SOVC on Reddit and is motivated by two goals: (1) providing researchers with an excellent resource for methodological decisions in studies of Reddit communities; and (2) creating the foundation for a new class of research methods and community support tools that reflect users' experiences of SOVC. To ensure that methods are respectfully and ethically designed in service and accountability to impacted communities, our work takes a qualitative, community-centered approach by engaging with two key stakeholder groups. First, we interviewed 21 researchers to understand how they study ``community'' on Reddit. Second, we surveyed 12 subreddits to gain insight into user experiences of SOVC. Results show that some research methods can broadly reflect users' SOVC regardless of the topic or type of subreddit. However, user responses also evidenced the existence of five distinct \textit{Community Archetypes}: Topical Q\&A, Learning \& Perspective Broadening, Social Support, Content Generation, and Affiliation with an Entity. We offer the Community Archetypes framework to support future work in designing methods that align more closely with user experiences of SOVC and to create community support tools that can meaningfully nourish the human need for SOC/SOVC in our modern world. 
\end{abstract}

%%
%% The code below is generated by the tool at http://dl.acm.org/ccs.cfm.
%% Please copy and paste the code instead of the example below.
%%

\begin{CCSXML}
<ccs2012>
<concept>
<concept_id>10003120.10003121.10011748</concept_id>
<concept_desc>Human-centered computing~Empirical studies in HCI</concept_desc>
<concept_significance>500</concept_significance>
</concept>
</ccs2012>
\end{CCSXML}

\ccsdesc[500]{Human-centered computing~Empirical studies in HCI}

\newcommand{\participants}{
\begin{table}[]
\begin{tabular}{lllll}
\toprule
\textbf{ID} & \textbf{Career Stage} & \textbf{Discipline} & \textbf{Methods} & %\# \textbf{Reddit Pubs} & 
\textbf{Reddit User?} \\ \midrule
R1 & Postdoc & CS, Math, Eng. & Quant %& 1--5
& prior to research \\ %\hline
R2 & PhD Student/Industry Researcher & CS, Math, Eng. & Mixed %& 1--5 
& prior to research \\ %\hline
R3* & Industry Researcher & CS, Math, Eng. & Qual %& \textgreater{}5 
& prior to research \\ %\hline
R4 & Faculty & CS, Math, Eng. & Mixed %& 1--5 
& prior to research \\ 
R5* & Faculty & CS, Math, Eng. & Qual %& \textgreater{}5 
& prior to research \\ %\hline
R6 & PhD Student & CS, Math, Eng. & Mixed %& 1--5 
& since research \\ %\hline
R7 & Industry Researcher & CS, Math, Eng. & Quant %& 1--5 
& prior to research \\ %\hline
R8 & Faculty & CS, Math, Eng. & Quant %& 1--5 
& never \\ 
R9 & Faculty & Humanities & Mixed %& 1--5 
& prior to research \\ %\hline
R10 & Faculty & Humanities & Qual %& 1--5 
& since research \\ %\hline
R11 & Faculty & Humanities & Qual %& 1--5 
& prior to research \\ %\hline
R12 & Faculty & Humanities & Mixed %& 1--5 
& prior to research \\ %\hline
R13 & Faculty & Humanities & Qual %& 1--5 
& never \\ 
R14 & Faculty & Medicine \& Health & Mixed %& 1--5 
& prior to research \\ %\hline
R15 & PhD Student/Industry Researcher & Medicine \& Health & Quant %& 1--5 
& prior to research \\
R16 & Faculty & Social Science & Qual %& 1--5 
& since research \\ %\hline
R17* & Faculty & Social Science & Qual %& \textgreater{}5 
& prior to research \\ %\hline
R18 & Faculty & Social Science & Mixed %& 1--5 
& prior to research \\ %\hline
R19 & Faculty & Social Science & Mixed %& 1--5 
& prior to research \\ %\hline
R20 & PhD Student & Social Science & Quant %& 1--5 
& prior to research \\ %\hline
R21 & Faculty & Social Science & Quant %& 1--5 
& prior to research \\ 
\bottomrule
\end{tabular}
\caption{Summary of Reddit Researcher Participants. Most participants indicated that they have published between 1--5 papers on Reddit; * indicates researchers who have published \textgreater{}5 papers on Reddit.}
\label{tab:participants}
\end{table}
}

\newcommand{\subreddits}{
\begin{table}[]
\footnotesize
% \begin{tabular}{l m{5.8cm} ll m{4cm}}
\begin{tabular}{lllll}
\toprule
\textbf{ID} & \textbf{Description} & \textbf{Size} & \textbf{$N$} & \textbf{Community Archetypes} \\ \midrule
S1 & Posts with niche absurdist humor style & 10-50K & 16 & ContentGen \\ %\hline
S2 & Corrections of poor scientific reporting & 10-50K & 28 & Learning \\ %\hline
S3 & Discuss fundamental elements of music & 100K-1M & 77 & Q\&A; Learning \\ %\hline
S4 & Help with job application materials & 100K-1M & 48 & Q\&A; ContentGen \\ %\hline
S5 & Support for pregnancy-related issues & \textless{}10k & 39 & Support \\ %\hline
S6 & POC/allies discuss race/intersectionality & 10-50K & 41 & Learning; Support \\ %\hline
S7 & Discuss ironic matters regarding race & \textless{}10k & 21 & Learning \\ %\hline
S8 & Support for a mental health issue & 50-100K & 13  & Support \\ %\hline
S9 & A US university and geographical area & 10-50K & 43  & Affiliation; Q\&A \\ %\hline
S10 & Q\&A with experts on an academic topic & 1-5+M & 25 & Q\&A; Learning \\ %\hline
S11 & Screenshots of social media posts from POC users & 1-5+M & 250 & ContentGen; Learning; Support \\ %\hline
S12 & Learning about indigenous spiritual beliefs & 50-100K & 7 & Learning; Support \\
\bottomrule
\end{tabular}
\caption{Descriptions of subreddits included in this study. %Portions of this table are reproduced verbatim from~\cite{smith_impact_2022} with permission from the authors. 
Abbreviations included in the Community Archetypes column are fully specified in Section~\ref{sec:communityarchetypes}; Archetypes are listed in order of highest to lowest relevance. Note: ``POC'' is an acronym for People of Color.}
\label{tab:subreddits}
\end{table}
}

\newcommand{\archetypes}{
\begin{table}[t]
\begin{tabular}{m{3.7cm} m{2.8cm} m{6.3cm}}
\toprule
\textbf{Archetype} & \textbf{Roles} & \textbf{Content Patterns} \\
\midrule
 \begin{tabular}{l} Sec.~\ref{sec:QA} \\ \textbf{Topical Q\&A} \\ (Q\&A) \end{tabular} & \begin{tabular}{l} \textbullet~Expert \\ \textbullet~Novice \end{tabular}
 &
 \begin{tabular}{m{6.2cm}} 
 \textbf{Posts:} questions  \\
 \textbf{Comments:} answers, discussions and enrichments of others' answers
 \end{tabular}  
 \\ \hline
 
 \begin{tabular}{l} Sec.~\ref{sec:learning} \\ \textbf{Learning \&} \\ \textbf{Broadening} \\ \textbf{Perspective} \\ (Learning) \end{tabular} & \begin{tabular}{l} \textbullet~Insider \\ \textbullet~Outsider \end{tabular} & 
  \begin{tabular}{m{6.2cm}} 
 \textbf{Posts:}  
news, events, publications, personal stories, questions on a particular topic \\
 \textbf{Comments:} celebrating or disagreeing, making jokes, contrasting or similar personal stories, elaborative ideas
 \end{tabular}  
 \\ \hline
 
 \begin{tabular}{l} Sec.~\ref{sec:support} \\ \textbf{Social Support} \\ (Support) \end{tabular} & \begin{tabular}{l} \textbullet~Support seeker %\\ \phantom{xx}(either for self, \\ \phantom{xx}or a loved one) 
 \\ \textbullet~Supporter \end{tabular} &
 \begin{tabular}{m{6.2cm}} \textbf{Posts:} personal experiences, sensitive disclosures, health questions, milestone announcements, reflection, venting, resources \\
 \textbf{Comments:} support, validation, reflection, commiseration, resources \end{tabular} \\ \hline
 
 \begin{tabular}{l} Sec.~\ref{sec:contentgen} \\ \textbf{Content} \\ \textbf{Generation} \\ (ContentGen) \end{tabular} & \begin{tabular}{l} \textbullet~Producer \\ \textbullet~Consumer \end{tabular} & 
  \begin{tabular}{m{6.2cm}} 
 \textbf{Posts:} original content from a sub member, or contributions from other users/platforms that exemplify a specific content style \\ 
 \textbf{Comments:} Opinions on the content, extra information, commiseration \\
 \end{tabular}  
 \\ \hline
 
 \begin{tabular}{l} Sec.~\ref{sec:affiliation} \\ \textbf{Affiliation with} \\ \textbf{an Entity} \\ (Affiliation) \end{tabular} & \begin{tabular}{l} \textbullet~Current affiliate \\ \textbullet~Prior affiliate \\ \textbullet~Future affiliate \end{tabular} &
  \begin{tabular}{m{6.2cm}} 
 \textbf{Posts:} entity-specific news, events, and questions \\
 \textbf{Comments:} Feelings about news or events, answers or advice about the entity
 \end{tabular}  
 \\ 
\bottomrule
\end{tabular}
\caption{Summary of Community Archetypes. We occasionally refer to these archetypes by the (parenthetical abbreviations) included in the first column.} %Note that Table~\ref{tab:subredditrules} provides additional information about the types of rules present in each community archetype.}
\label{tab:archetypes}
\end{table}
}

\newcommand{\constructs}{
\begin{table}[t]
\begin{tabular}{ll}
\toprule
\textbf{Construct} & \textbf{Assessment Strategies} \\
\midrule
Interactivity (sec.~\ref{sec:interactivity})& \begin{tabular}[c]{@{}l@{}}\textbullet~Volume of observable interaction\\ \textbullet~Degree of conversation occurring\end{tabular} \\ \hline

Membership Boundaries (sec.~\ref{sec:membership}) & \begin{tabular}[c]{@{}l@{}}\textbullet~Tiers of membership\\\textbullet~Prolonged participation\\ \textbullet~Linguistic membership boundaries\end{tabular} \\ \hline

Homogeneity (sec.~\ref{sec:homo}) & \begin{tabular}[c]{@{}l@{}}
\textbullet~Demographic\\ \textbullet~Situational \\ \textbullet~Goals or values \\ \textbullet~Linguistic similarity\end{tabular} \\ \hline

Norm Enforcement (sec.~\ref{sec:norms}) & \begin{tabular}[c]{@{}l@{}}\textbullet~Vertical moderation (by moderators)\\ \textbullet~Horizontal norm enforcement (by users)
\end{tabular} \\
\bottomrule
\end{tabular}
\caption{Assessment strategies for antecedent constructs to sense of virtual community.}
\label{tab:constructs}
\end{table}
}

\newcommand{\researchapproach}{
\begin{figure}
    \centering
    %\frame{
    \includegraphics[width=\textwidth]{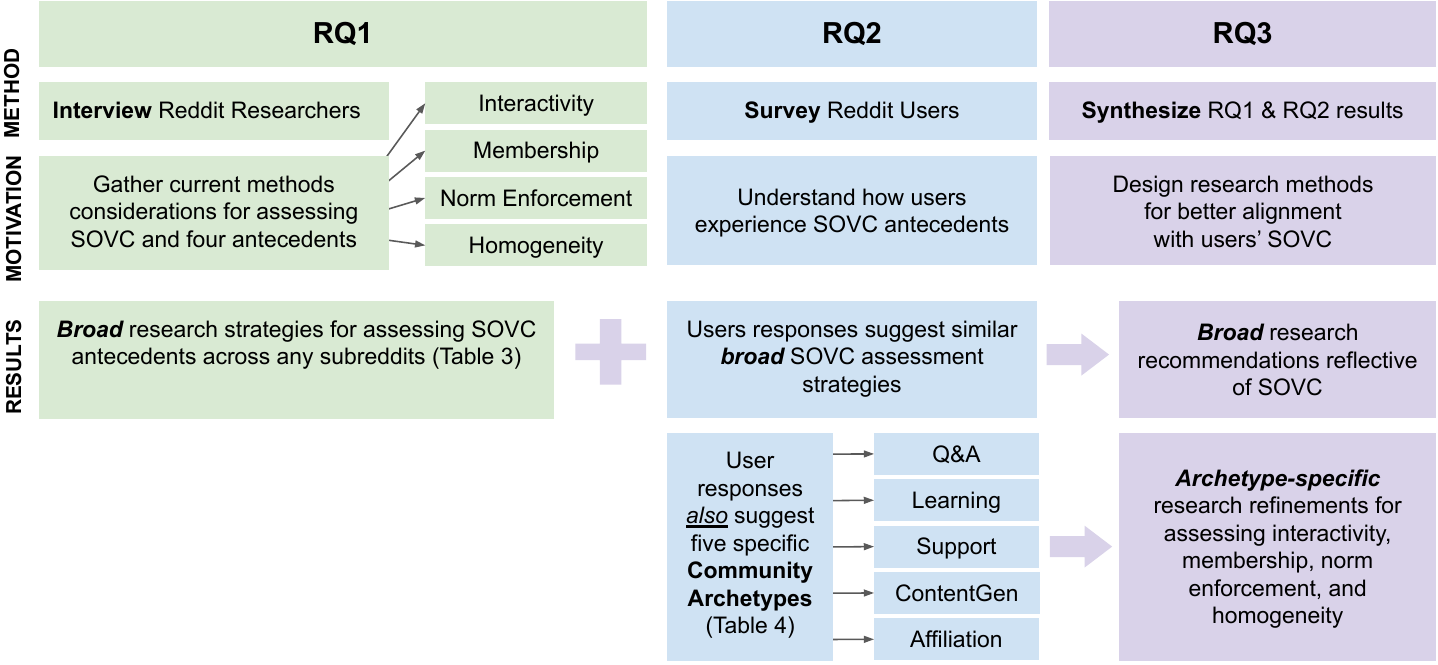}%}
    \caption{Summary of Research Approach.}
    \label{fig:researchapproach}
\end{figure}

}

\newcommand{\subredditrules}{

\begin{table}[]
\begin{tabular}{l | lllll}
\toprule
                                    & \textbf{Q\&A}     & \textbf{Learning}     & \textbf{Support}     & \textbf{Content}     & \textbf{Affiliation}    \\ \midrule
Total \# of Subs in Sample          & 4        & 7            & 5           & 3              & 1      \\
Total \# Unique Rules               & 28       & 39           & 47          & 24             & 10     \\
Average \# Rules/Sub                 & 7        & 5.6          & 9.4         & 8              & -      \\ \midrule \midrule
Rule Type                           & \multicolumn{5}{c}{\% of of total rules per archetype category} \\ \midrule

Restrictive (46.90\% across \textit{all} Reddit)                        & 55.2\%     & 52.5\%         & 60\%          & 50\%             & 70\%     \\
Prescriptive (39.35\%)                       & 44.8\%     & 47.5\%         & 40\% %.4\%       
& 50\%             & 20\%     \\ \hline

Advertising/Commercialization (5.49\%)   & 10.3\%     & 7.5\%          & 8.5\%         & 12.5\%           & 10\%     \\
Consequences/Moderation (20.80\%) & 3.4\%      & 5\%           & 4.3\%         & 4.2\%            & 10\%     \\
Content/Behavior (71.95\%)                    & 55.2\%     & 62.5\%         & 61.7\%        & 62.5\%           & 40\%     \\
Doxxing/Personal Info (2.88\%)               & 7\%        & 2.5\%          & 2.1\%         & 8.3\%            & 10\%     \\
Format (13.96\%)                              & -        & 2.5\%          & -           & -              & -      \\
Harassment (4.14\%)                          & 7\%        & 15\%           & 6.4\%         & 8.3\%            & 10\%     \\
Hate Speech (3.22\%)                         & -        & 15\%           & 10.6\%        & 12.5\%           & -      \\
Links/Outside Content (9.71\%)            & 3.4\%      & 2.5\%          & 2.1         &        -        & -      \\
Low-Quality Content (2.38\%)                 & 10.3\%     & 10\%           & 4.2\%         & 16.7\%           & -      \\
NSFW (4.62\%)                               & 3.4\%      & -            & 2.1\%         & 4.2\%            & -      \\
Off-Topic (7.23\%)                           & 10.3\%     & 7.5\%          & 14.9\%        & 8.3\%            & 10\%     \\
Personality (6.44\%)                        & 3.4\%     & 5\%            & 4.2\%         &         -       & -      \\
Reddiquette (3.75\%)                         & 3.4\%      & 2.5\%          & -           &      -          & -      \\
Reposting (2.32\%)                           & 3.4\%      & 2.5\%          & 2.1\%         & 4.2\%            & 10\%     \\
Spam (3.85\%)                               & 3.4\%      & 2.5\%          & 4.2\%         & 4.2\%            & -      \\
Spoilers (1.74\%)                           & -        & -            & 4.2\%         & -              & -      \\
Trolling (1.56\%)                           & -        & 10\%           & 8.5\%         & 4.2\%            & -     

\\ \bottomrule

\end{tabular}
\caption{Summary of Rules Analysis. For context and comparison, we parenthetically include the percentage of these rule categories across a large random sample of manually-coded subreddits, as reported in~\cite{fiesler_reddit_2018}. Note that 5 categories of rules did not appear in our data set.}
\label{tab:subredditrules}
\end{table}
}

\newcommand{\UIcomparison}{
\begin{figure}
    \centering
    \frame{\includegraphics[width=\textwidth]{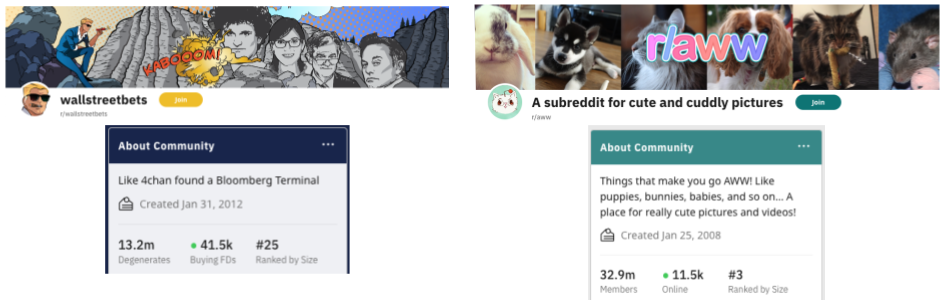}}
    \caption{Screenshots of banner images and community descriptions at \texttt{r/wallstreetbets} (left) and \texttt{r/aww} (right). Retrieved in December 2022.}
    \label{fig:UIcomparison}

    \frame{\includegraphics[width=0.4\textwidth]{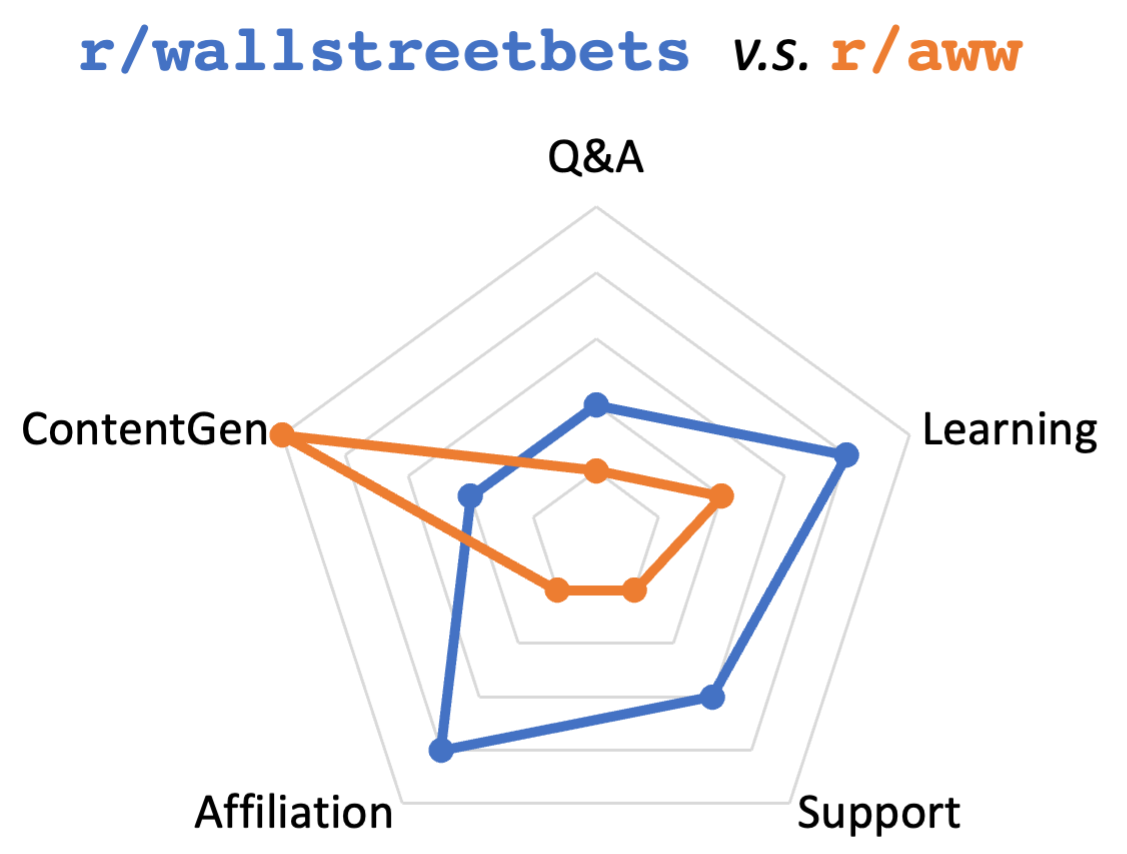}}
    \caption{Hypothetical spiderplot comparing Community Archetypes in \textcolor{blue}{\texttt{r/wallstreetbets} (blue)} \textit{v.s.} \textcolor{orange}{\texttt{r/aww} (orange)}  for purposes of illustration only. Rather than factual graphs, these plots are suggestive illustrations based on our estimated observations of the proportional frequency of visible content patterns and user behaviors. The outermost ring represents 100\% frequency, with concentric rings descending to an innermost ring of near 0\% frequency. Future work could systematically calculate such plots by tabulating the actual frequency of posts and user behaviors in each archetype.}
    \label{fig:comparison}
\end{figure}
}

\keywords{Sense of community, sense of virtual community, online communities, community support, Reddit, survey, interview, data science, users, researchers}
\maketitle

%\newpage
\input{1_intro}

\input{2_background}
\input{3_methods}
\input{4_results_RQ1}

\input{5_results_RQ2}

\input{6_discussion}
\begin{acks}
%Acknowledgements omitted for review.
We thank our participants for sharing their insights and experiences. We are also appreciative for the supportive feedback from Stevie Chancellor, Loren Terveen, Charles Chuankai Zhang, Sanjay Kairam, and the dedicated anonymous reviewers who supported us to improve this manuscript. Undergraduate assistants Tamer Shahwan, Sahand Setareh, and Vikki Wong helped us to collect our survey data and were also instrumental to the success of this project. This work was supported by NSF Award \#1910225.
\end{acks}

\bibliographystyle{ACM-Reference-Format}
\bibliography{SOVC_Reddit}

\appendix
\newpage
\input{interviewprotocol}

\end{document}

%% file: 1_intro.tex
\section{Introduction}

%SUMMARY OF REVISIONS DOCUMENT: \url{https://docs.google.com/document/d/1oATA_TH6NW7nOdBw5xxCauIQLBDo5N6tSM48409HIdM/edit}.

%SUPPLEMENTAL MATERIALS:
% \url{https://www.overleaf.com/project/639cdeed05c59de3fddce9fb}

For decades, concerning trends point toward the degradation of in-person communities and a growing epidemic of loneliness~\cite{twenge_less_2019}. Concurrently, participation in online communities has exploded, with rapidly evolving moderation affordances~\cite{bajpai_harmonizing_2022} and styles~\cite{seering_metaphors_2022}, rules and norms~\cite{fiesler_reddit_2018,frey_governing_2022,chandrasekharan_internets_2018}, and user groups who may or may not be connected offline~\cite{zhang_integrating_2011,smith_i_2020}. For example, people might not talk to neighbors like they used to, but they might also go on Nextdoor or their city's subreddit to observe neighbors' discussions. Academics might opt to work from home more frequently and miss out on water cooler chat, but check Twitter twice an hour to catch the latest hot takes. People suffering from stigmatized illness may never speak a word of it at the office, later going home to an intimate Zoom chat hosted by a private Facebook group for patients. 

Examples like these make it clear that how modern humans participate in communities---and consequently, how we experience a Sense of Community (SOC) or Sense of Virtual Community (SOVC)---have shifted over time across different geographical, cultural, and sociotechnical contexts. Studying how people experience SOC/SOVC is meaningful because it is a deep and unavoidable human need. The access to or lack of community has profound psychological and material impacts on people's lives. For example, SOC impacts life satisfaction~\cite{prezza_sense_1998}, well-being~\cite{chavis_meaning_1986}, perceived safety~\cite{perkins_ecological_1996}, problem solving~\cite{bachrach_coping_1985}, and social or political participation~\cite{obst_exploration_2002}. 
%Higher SOC can increase in-group cohesion and positive self-image~\cite{tajfel_human_1981}, yet it can also contribute to a negative collective identity in segregated, minoritized, or stigmatized communities~\cite{fisher_psychological_2002}. 
%As one case-in-point alluded to above, online health spaces can provide support, resources, and connection to other users, especially when they cannot find it elsewhere (e.g., due to COVID-19~\cite{meurer_digital_2022} or geographically sparse patient populations~\cite{sheldon_online_2021, walsh_qualitative_2021, shuter_randomized_2022}). Yet online spaces can also expose users to misinformation, stigmatizing comments, or even witch hunts~\cite{bbc_reddit_2013} and harassment~\cite{mortensen_anger_2018}. 
Considerations like these underpin the importance of improving our scientific understanding of: how we develop SOC/SOVC; %techniques for understanding when communities are strong and healthy; 
how we can best understand, measure, and support the formation of healthy communities; and how SOC/SOVC can either nourish our needs or cause damage, depending on the circumstances.

This paper investigates SOVC on Reddit, a platform that serves millions of users and communities worldwide and has attracted growing research attention~\cite{proferes_studying_2021}.\footnote{This paper assumes familiarity with Reddit-specific terminologies, however supplemental material section 1 includes detailed descriptions of all terms, such as ``subreddit,'' ``thread,'' ``Original Poster'' (OP), etc. for reference.} Our work is motivated by two overarching goals: (1) to provide future researchers with an excellent resource for thinking through major methodological decisions in studies of Reddit communities; and (2) to provide the foundation for a new class of data science techniques and data-driven community support tools. Regarding the first goal, studies that aim to foster healthy online communities should ensure that community spaces and members are properly identified, and that researchers' selected methods align with community members' experiences and values. Although much prior work refers to each subreddit as its own community, recent work suggests that individual subreddits are not the best ``units'' of community. Most users report greater SOVC across \textit{multiple} subreddits~\cite{smith_impact_2022}, and each subreddit in these \textit{sets} may meet different needs~\cite{hwang_why_2021,teblunthuis_identifying_2022}. If researchers assume that each subreddit is individually experienced as a community, they risk inaccurate conclusions, or they may miss key aspects of the larger picture of user experience. Providing better ways to identify community spaces will enable researchers to better support the human need for SOC.

\researchapproach

Drawing upon prior work in organizational psychology and HCI, our second goal aims to advance our ability to measure and predict SOVC. Inspired by recent work on Twitch~\cite{kairam_social-ecological_2022}, we are motivated by the promising opportunity to design and evaluate human-centered data science techniques. However, academic constructions of SOC/SOVC do not always align with the lived experiences of community members~\cite{pretty_sense_1996,puddifoot_dimensions_1995,rapley_playing_1999}. Consequently, ``\textit{the problems raised by the use of the notion of community are increased if the gap between academic and lay meaning is not taken into account.}''~\cite{mannarini_multiple_2009} To reduce this gap, we took an empirical, qualitative, and community-centered approach to developing a guiding framework for new methods. We engaged with two stakeholder groups--Reddit researchers \textit{and} Reddit users--to ask:

\begin{description}
\item[RQ1:] How do researchers conceptualize and operationalize virtual community on Reddit?% in their work?
\item[RQ2:] How do Reddit users experience a sense of virtual community in their use of the platform? And how does this vary across communities?
\item[RQ3:] How can we operationalize important community-related concepts to better align researchers' methods with users' experiences? 
\end{description}

Figure~\ref{fig:researchapproach} overviews our research approach. We conducted interviews with 21 researchers to address RQ1 and found that they typically view subreddits as topical affinity groups; not all subreddits can or should be described as communities. Rather, certain types of interactivity, membership boundaries, homogeneity, and norm enforcement may distinguish non-community affinity groups from communities (Table~\ref{tab:constructs}). For RQ2, we surveyed 12 subreddits. User responses confirmed that their SOVC formation is broadly tied to how researchers tend to study communities. However, users expect nuanced forms of community activity in different types of subreddits; their experiences of SOVC arise primarily when these specialized activity forms are happening well and frequently. Thus our data suggest the existence of five underlying \textbf{\textit{Community Archetypes}} distinguished by particular user roles and content patterns (Table~\ref{tab:archetypes}). Our discussion synthesizes these two sets of results to address RQ3. Some methods for measuring community structures and SOC/SOVC can be used broadly, no matter the type of subreddit, whereas others must be designed according to which archetype(s) the subreddit(s) embody. We provide examples to illustrate these concepts within existing subreddits and studies. Finally, we offer a road map for future work to apply the Community Archetypes framework and to create community support tools that can meaningfully nourish the human need for SOC/SOVC in our modern world.

%% file: 2_background.tex
\section{Related Literature}
\subsection{The history of sense of community research}

\paragraph{Organizational and community psychology}
Long before the emergence of the Internet, researchers in organizational and community psychology devised varied ways of assessing the structural elements that define geographical communities, and for measuring sense of community (SOC), examining factors like common fate~\cite{campbell_common_1958}, supportive climate~\cite{doolittle_communication_1978}, and the amount of time one expects to stay~\cite{glynn_psychological_1981,gusfield_community_1975}. As research advanced, definitions became less attached to geographical centers and more inclusive of \textit{relational} communities denoted by facets like interpersonal connection, interests, and hobbies~\cite{gusfield_community_1975, mcmillan_sense_1986, ashforth_social_1989, lickel_elements_2001}. One prominent theory of SOC by McMillan and Chavis (1986) proposes four elements: membership, influence, fulfillment of needs, and shared emotional connection~\cite{mcmillan_sense_1986}. This theory can be applied to geographical \textit{or} relational communities and remains popular today, with occasional uses in HCI (\textit{e.g.,}~\cite{kim_designing_2022}). 

\paragraph{Sociology and communication studies}
The fields of sociology and communication studies have also embraced a relational conception of communities. For example, Benedict Anderson examined national identity as a community builder. Although all citizens do not know each other personally, they may nonetheless view themselves as members in the ``imagined community'' of their nation~\cite{anderson_benedict_imagined_1983}. This concept of imagined communities provides a helpful lens for online community scholars to understand how users can build communities online without either geographical closeness or personal connections. Sociologists in the 90s have also examined the weak and strong ties that users made both off- and online and found that ties still existed online, despite users not having any geographical proximity~\cite{wellman_barry_net_1999}. Communications scholars have also operated with the idea that communication is a major indicator and strengthener of communities both off- and online since at least the 90s~\cite{rothenbuhler_communication_1996, shah_communication_2001}. Social media may simply be the next step in humans' millennia-long history in mediated, text-based communities rather than a fundamentally ``new'' innovation in the course of human history~\cite{standage_tom_writing_2015}.

\paragraph{Measuring SOC/SOVC}
While decades of scholarship across different academic disciplines grappled with what a community is (as does our present work in this paper), people seem to reliably know when they are a part of communities and when they are not~\cite{lickel_elements_2001,sarason_psychological_1974, anderson_imagined_2006}, allowing researchers to use their cognitive evaluations to measure SOC using psychometric instrumentation~\cite{hoyle_handbook_2013}. While early scales were developed for offline communities, researchers in the early 2000s began to investigate how online communities may differ from offline and to create new psychometric measures for SOVC~\cite{koh_sense_2003,blanchard_experienced_2004}.\footnote{For one example, people in online communities are not embodied but are represented by online profiles; consequently issues of purposefully creating an identity, and making identifications of others, are possibly more important to SOVC, relative to in-person communities, where much more information is naturally communicated about identity by physical appearance and proximity~\cite{blanchard_experienced_2004}.} One critical issue with these preliminary scales is that they conflated the cognitive evaluation of SOVC with a variety of psychological factors that cause SOVC to develop. Therefore, a more modern approach seeks to distinguish \textit{antecedents} to SOVC from the \textit{experience of SOVC} itself, as well as from specific \textit{outcomes} of SOVC. For example, similarity, interactivity, membership boundaries, common goals, and history of interactions can be distinguished as individual psychological constructs which all \textit{precede} SOVC. Similarly, identification with the group, commitment to the group, group satisfaction, and centrality are all distinct, measurable psychological outcomes, when SOVC exists~\cite{blanchard_developing_2020}. Because of the increasing prevalence of automated bots as social actors in online communities~\cite{seering_it_2020,seering_social_2018}, recent work also introduced bot governance as a new SOVC antecedent~\cite{smith_impact_2022}.

\paragraph{Analogizing aspects of behavioral trace data as SOVC antecedents}
Our work's theoretical basis benefits from the recent move in organizational psychology to distinguish antecedents from SOVC and its outcomes. In particular, SOVC is an internal and highly variable user experience; it cannot be directly measured via behavioral trace data. However, we aim to understand how users' internal experiences of \textit{antecedents} to SOVC may relate to empirical measures that \textit{can} be calculated from behavioral trace data. If we can develop data science-based analogs of known SOVC antecedents, we may be able to effectively predict SOVC. As described in section~\ref{sec:antecedents}, this work uses four of the most well-studied SOVC antecedents to structure our interviews with researchers: interactivity, homogeneity, norm enforcement, and membership boundaries.

\subsection{Bridging from organizational psychology to HCI}
HCI and social computing have a long and rich history of studying online communities. However, this history has largely developed adjacent to the literature in organizational psychology. Studies of platforms like Facebook~\cite{zhang_integrating_2011}, YouTube~\cite{rotman_community_2009}, Twitter~\cite{blight_sense_2017,ricoy_twitter_2016,java_why_2007}, Twitch~\cite{kairam_social-ecological_2022}, Discord~\cite{jiang_moderation_2019}, Instagram~\cite{blight_sense_2017,thomas_instagram_2020}, TikTok~\cite{bonifazi_investigating_2022}, and Wikipedia~\cite{baytiyeh_volunteers_2010,pentzold_imagining_2011} have explored whether and how users experience community and sense of community on the platform. However, other papers use ``community'' without specifically justifying it. Prior work points to the importance of triangulating qualitative and quantitative data to develop deeper understandings of online communities~\cite{rotman_community_2009,kairam_social-ecological_2022}. As we aim to develop more precise measures and techniques, incorporating concepts across organizational psychology, HCI, and users' subjective perspectives will improve our ability to delineate and study what users experience as community and how to build community support tools---\textit{e.g.}, moderation policies and toolkits~\cite{zhang_policykit_2020,schneider_modular_2021}, community-in-the-loop algorithms~\cite{smith_keeping_2020} and governance bots~\cite{smith_impact_2022}. Therefore, we will next summarize four main ways that the HCI literature has previously delineated online communities: platform affordances; shared interests; common vernacular; and interactivity. 

\textbf{Platform affordances.} Researchers often designate communities by using particular platform affordances. For example, social network analysis analyzes connections between users who have ``added'' or ``followed'' each other. Users are considered ``nodes'' or ``vertices'' in the network, connections between them are ``edges,'' and researchers detect communities based on communication patterns, edge weights, \textit{etc}.~\cite{fortunato_community_2016} Some studies use hashtags to indicate communities, such as Black Twitter~\cite{klassen_more_2021,sharma_black_2013} or communities of users on Tumbler~\cite{griffith_behind_2021} or Instagram~\cite{andalibi_sensitive_2017,mcewan_communication_2016} who use the hashtag \#depression. On other platforms, communities are designated by more concrete digital ``containers'' that users elect to join, \textit{e.g.}, Facebook Groups~\cite{zhang_facebook_2013}, online health blogs~\cite{smith_i_2020,levonian_patterns_2021}, or Twitch channels~\cite{hamilton_streaming_2014}. In line with this ``container'' view, Reddit's website copy and many papers (\textit{e.g.},~\cite{weld_making_2021, tan_tracing_2018, cagle_shades_2019, kou_supporting_2017, prakasam_reddit_2021, an_political_2019, potts_boycotting_2019}) refer to each subreddit as its own community. Other papers don't mention community (\textit{e.g.},~\cite{hermes_hating_2019, kou_conspiracy_2017}) or identify families of subreddits~\cite{tan_tracing_2018} with overlapping membership~\cite{teblunthuis_identifying_2022}. Moreover, most users do not perceive individual subreddits as the best ``units'' of community, instead reporting greater SOVC across \textit{multiple} subreddits~\cite{smith_impact_2022}, each of which may meet different needs~\cite{hwang_why_2021}. These studies suggest that users' SOVC often extends beyond the boundaries of digital containers, and that research will benefit from more fine-grained delineations. %Other HCI perspectives, however, do account for online communities beyond platform affordances.

\textbf{Shared interests.} Classic works in HCI denote online community as a group of users with a shared interest or goal~\cite{preece_online_2000,bieber_toward_2002,kraut_building_2012,mcewan_communication_2016}. If users simply exist in the same online space with nothing in common, it is unlikely that they will form community-like connections. However, when users have a common interest or goal driving them together, community bonds are much more likely to form. Shared interests can be such a strongly unifying force that users may attempt to migrate their communities to new platforms when current platforms are insufficient or threatened. For example, communities of fandom users have often migrated across platforms~\cite{fiesler_moving_2020} and hundreds of thousands of Twitter users rapidly migrated to Mastodon after the 2022 Musk acquisition~\cite{keegan_what_2022}.

\textbf{Common vernacular.} Another approach from natural language processing that is commonly used in HCI focuses on the common vernacular that users employ when interacting in online spaces. Use and familiarity with specific language gives users the sense that they are part of an ``in-group.''~\cite{johnstone_discourse_2008} Therefore, the more that specific forms of language, discursive conventions, or terminologies exist in an online space, the more likely it is to be a community~\cite{johnstone_discourse_2008,danescu-niculescu-mizil_no_2013}.

\textbf{Interactivity.} Finally, HCI researchers have focused broadly on interactivity: the more activity that exists in a group of users, the stronger the community~\cite{preece_online_2000, mcewan_communication_2016, danescu-niculescu-mizil_no_2013}. Some HCI work considers interactivity as a mechanism for designating user groups, such as core \textit{v.s.} peripheral community members~\cite{bryant_becoming_2005,halfaker_making_2013}, old-timers \textit{v.s.} newcomers~\cite{choi_socialization_2010,yang_commitment_2017}, or differing user roles~\cite{zhang_working_2022,muller-birn_peer-production_2015,jin_beyond_2007}. Diverging from considerations of interactivity as a determination of membership or as an \textit{antecedent} to SOVC however, much work refers to the type and volume of interaction instead as a \textit{success} metric. For example, \citeauthor{cunha_are_2019} describe four success measures: (1)~growth in the number of members; (2)~retention of members; (3)~long-term survival of the community; (4)~volume of activity~\cite{cunha_are_2019}.
Many studies take a similar approach, with community size or survival as major targets of prediction. Recent work has also sought more refined metrics, such as measuring \textit{pro-social} behaviors, rather than identifying \textit{all} interaction as positive, community-oriented behavior~\cite{bao_conversations_2021}. This prior work provides important insights, however such HCI methods can be refined through: (1) a more coherent alignment with advances in organizational psychology theory that distinguish between antecedents, SOVC, and outcomes of SOVC; and (2) a community-centered approach that reflects users' actual experiences of SOVC in the design of research methods. These considerations motivate our research questions and goal of creating an empirical framework to guide future community data science techniques.%, and our selection of the three research questions in our introduction.

%% file: 3_methods.tex
\section{Methods}
Figure~\ref{fig:researchapproach} summarizes our research approach. We interviewed 21 researchers to understand how they conceptualize and operationalize community on Reddit. We also observed and surveyed 12 different subreddits to gain insight into users' experiences of SOVC. We then used Grounded Theory Method~\cite{muller_curiosity_2014} to analyze the data. This study was reviewed and deemed exempt by our institutional IRB office. We begin with a positionality statement to disclose how our own personal identities have influenced our study before describing our methods in detail.

\subsection{Positionality \& Ethical Stance}\label{sec:positionality}
We view the increasing research attention on Reddit~\cite{proferes_studying_2021} as an urgent opportunity to ensure that new methods are conscientiously, respectfully, and ethically designed for purposes of service and accountability to communities rather than extraction or manipulation. All authors are regular Reddit users.\footnote{Three team members, including the first author and two assistants listed in the acknowledgements, were originally recruited to join this research project through a subreddit post.} Consequently, we are influenced by our own on- and offline experiences related to our usage of Reddit. Most subreddits were selected somewhat randomly according to criteria we will soon describe. However, our own membership and prior engagement with moderators contributed to our ability to successfully recruit a few of the subreddits in Table~\ref{tab:subreddits}. For example, the second author is a member of S8. Due to research and personal interests in mental health, she felt it vital to include at least one mental health subreddit. Given an otherwise arbitrary set of candidates, it was most sensible to recruit a familiar subreddit. As academic researchers with values and ethics oriented toward equity and inclusion, we also felt it was vital to recruit subreddits for underrepresented populations (\textit{e.g.,} S6, S7, S11), even though we do not identify as members. %Many users who belong to vulnerable and/or marginalized populations derive an important source of SOC/SOVC from Reddit. Therefore,
We acknowledge that a complex set of privacy, safety, and equity concerns must be addressed in the development of new methods that can measure and impact human communities. Our research team values collaborative efforts with moderation teams and users, and we suggest that future researchers continue to work directly with users and communities, especially ensuring that such research efforts are allowed and desired. %Given these considerations, we next describe our methods in more detail. 

\subsection{Interviews with Reddit Researchers}\label{sec:interviews}
During our literature review, we observed that most papers on Reddit seem to be based on unstated assumptions about subreddits existing as communities. However, literature review does not enable us to make claims about Reddit researchers’ actual mental models of community, how these impact selected methodologies, or other unwritten limitations or opportunities they have considered. We chose to interview Reddit researchers in order to ensure the empirical validity and rigor of our claims and to directly compare researcher and user perspectives.

\subsubsection{Recruitment}
A systematic review of 727 Reddit publications from 2010--2020 shows that 57.4\% are first-authored by researchers in Computer Science, Math, and Engineering units, with disciplines like the Humanities (18.2\%), Medicine \& Health (9.6\%), and Social Science (7.8\%) accounting for most of the rest. 482 (66.3\%) papers used quantitative methods (primarily computational techniques such as machine learning and natural language processing), 183 (25.2\%) used qualitative methods (\textit{e.g.}, discourse analysis, ethnography, grounded theory), and 56 (7.7\%) used mixed methods~\cite{proferes_studying_2021}. We manually collated a list of email addresses for all first and last authors in this set of papers, on the assumption that first and last authors are most influential in methods decisions. We emailed the list a recruitment message and screening survey and selected researchers across disciplines, methods, and seniority, purposefully ensuring representation of as balanced and diverse a collection of voices as possible. We continued recruiting until we reached data saturation--\textit{i.e.} new interviews were no longer providing new information~\cite{muller_curiosity_2014}. %Because we recruit from and report results back to the community being studied, obscuring the identity of our participants faces a \textit{small population} challenge~\cite{saunders_anonymising_2015}. We do not require an extreme technique such as ``un-Googling''~\cite{shklovski_-googling_2013}, since we do not consider researchers a vulnerable population. Adhering to best practices~\cite{kaiser_protecting_2009,lee_lets_2013}, we use participant IDs, and do not connect those IDs to names or demographic information. 
Table~\ref{tab:participants} summarizes participant information of 21 participants: 13 faculty, 1 postdoc, 2 industry researchers, and 4 Ph.D. students (2 of whom also indicated industry researcher). 7 participants (33.3\%) self-identified Computer Science, Math, and Engineering (including HCI) as their primary discipline, 6 (28.6\%) Social Science, 5 (23.4\%) Humanities, and 2 (9.5\%) Medicine \& Health. Of these, 6 primarily used quantitative methods, 7 qualitative, and 8 mixed. 14 participants (66.7\%) identified as white, 6 as Asian (28.6\%), and 1 as Middle Eastern. 16 participants (76.2\%) were at a US institution, 4 in Europe (19\%), and 1 in Asia. 13 identified as male (including 1 trans) (61.9\%), and 8 female (38.1\%).

\participants

\subsubsection{Interview Protocol}\label{sec:interviewprotocol}
Interviews were conducted remotely over Zoom by the two co-first authors. Interviews ranged from 43 to 73 minutes long and lasted an average length of 56 minutes. Interviews had two parts. The first used a semi-structured approach with questions regarding the researchers' mental model of community, what a strong community might look like, specific definitions/operationalizations of community, and methods decisions in the interviewees' work. (The complete set of interview questions is available in supplemental section 3.) The second part used a talk-aloud structure (similar to~\cite{fisler_sometimes_2017, sripada_structure_2020}) in which we screen-shared slides, asking participants to reflect on four known antecedents to SOVC (interactivity, membership boundaries, homogeneity, and norm enforcement). We defined each antecedent on-screen and asked how researchers might measure it using any qualitative, quantitative, or mixed methods. A final slide asked about any additional variables or ideas we had not asked about. Next, we describe why we chose these four particular antecedents.

 %With the definitions as a visual reference point, we asked questions about each variable in their area of research. For example, a quantitative researcher would be asked to provide specific parameters and operationalization for the variables. In contrast, qualitative researchers would be asked how each variable could affect community and how they might determine if it were present.  

\subsubsection{Selected antecedents to sense of virtual community}\label{sec:antecedents}
Organizational psychologists have demonstrated interactivity, membership boundaries, homogeneity, and norm enforcement to be fundamental antecedents to the formation of SOVC~\cite{blanchard_developing_2020}. This list is not exhaustive, however, it does provide a strong theoretical bedrock.\footnote{See supplemental section 2 for in-depth discussion of the theoretical bases for these antecedents.} We opted for more generic rather than overly-specified constructs to allow for the broadest possible interpretations of user behavior. For example, we asked about \textit{interactivity} rather than \textit{support exchange} or \textit{information sharing}, since both of those are more specific forms of interactivity which may or may not be \textit{required} for SOVC in different virtual contexts. We displayed the following four definitions to participants:

\begin{description}
\item[\textbf{Interactivity:}] The degree to which users are interacting within the community.
\item[\textbf{Membership Boundaries:}] The degree to which users can be considered members of a community, \textit{v.s.} not being considered members of a community.
\item[\textbf{Homogeneity:}] The degree of similarity of users related to their values, attitudes, goals, or other personal characteristics (\textit{e.g.}, demographics, personality traits, etc.).
\item[\textbf{Norm Enforcement:}] The degree to which norms are enforced within the community.
\end{description}

\subsubsection{Interview limitations}
One limitation is that we could not interview researchers who directly studied the \textit{same} subreddits we surveyed, however, we addressed this limitation by ensuring that diverse subreddits were included in our surveys, including the same basic types studied by prior researchers. Standard interview limitations apply: (1) Our small sample of Reddit researchers may not be representative of all researchers; and (2) We asked researchers to describe prior work, some of which was completed years ago, and recall of past events is imperfect.

\subsection{Surveys of Reddit Users}

\subreddits

Surveys were run in collaboration with the moderation teams of twelve different subreddits (See Table~\ref{tab:subreddits}).\footnote{This paper uses the same survey data collected in~\cite{smith_impact_2022}, which performs a purely statistical analysis on Likert rating data for psychometric scale validation. The present paper uses and analyzes qualitative data from the surveys in order to understand new dimensions of sense of virtual community. Complete survey text is available at \url{https://bit.ly/gov-bots_pdf}.} 
In addition to the reasons mentioned above in sec.~\ref{sec:positionality}), we selected subreddits using the following criteria. We avoided NSFW (Not Safe For Work) and hate- or abuse-centered subreddits for two main reasons: (1)~preserving the safety of our research team; (2)~focusing our research efforts on \textit{pro-social} communities---\textit{i.e.}, the types of communities that we hope this work can support. We aimed to capture as diverse a collection of subreddits as possible, with high variability in size, topic, and observable community activities (\textit{e.g.}, weekly or daily threads, annual customs, special AMA events, different types of user flairs or threads, etc.) to ensure that our results are as broadly applicable across Reddit as possible. We also chose \textit{active} subreddits in which surveys were allowed and in which users appear to post every day (rather than those with obviously inactive userbases) to maximize response rates. We posted surveys to each subreddit using a method preferred by each moderation team, including strategies such as: 

\begin{itemize}
    \item Moderators create a post about the survey and pin it to the top of the subreddit.
    \item Researchers post the survey in a weekly thread for surveys and/or personal promotion (and mods ``approve'' or highlight the post).
    \item Researchers post the survey to a thread specifically for surveys in that subreddit.
\end{itemize}

Survey links were posted from October through December of 2021, and each link was open to responses for a maximum of two weeks; some surveys were closed before that time due to satisfactory response rates. Survey participants could provide their email to opt into a drawing for \$10 eGift Cards. In addition to basic demographics, surveys included three free response questions that provide insight into users' experience of SOVC with an eye toward RQ2 and RQ3:

\begin{itemize}
    \item What motivates you to visit \texttt{r/[subreddit name]}?
    \item What makes \texttt{r/[subreddit name]} feel like a community to you?
    \item What (if anything) could be done to improve your experience in \texttt{r/[subreddit name]}?
\end{itemize}

\subsubsection{Survey participants}\label{sec:demo}
We manually inspected all survey responses and eliminated those which contained nonsensical or copy/pasted free responses. In total, we collected 608 valid responses across the twelve subreddits in Table~\ref{tab:subreddits}. Respondents skew male (58.8\%), white (61.5\%), residing in North America (81\%), and relatively young (50.1\% of respondents selected 25--34 years of age, and another 21.3\% 18--24); these demographic trends are broadly consistent across the Reddit userbase. 

 %were , 18.8\% were 35--44, and 7.5\% were 45--54.  respondents identified as only White or Caucasian, 15.1\% as only Black or African American, 4.8\% as only Asian or Pacific Islander, 3.6\% as only Hispanic, and less than 3\% in other categories, including multiple races. 41.7\% of respondents reported having a bachelor's degree, 20.8\% a post-bachelor's degree, 19.6\% a high school diploma or GED, and 16.3\% a trade or associate degree. 73.9\% of respondents reported being employed, 14.5\% reported being students, and 6.0\% reported being unemployed. 81\% of respondents reported living in North America, 7.7\% in Europe, 6.6\% in North America outside the U.S., and less than 3\% elsewhere. 50.8\% of respondents reported using their sub for 1--5 years, 20.9\% 6 months--1 year, and 12.8\% 1--5 months. 41.4\% of respondents report visiting their sub ``almost daily'', 27.3\% about once per week, and 20.4\% multiple times per day.

\subsubsection{Survey limitations}
The most important survey limitation lies in our selection of subreddits. Since every subreddit can have unique rules, norms, interaction styles, bots, \textit{etc}., it is impossible to select a perfectly representative set.For instance, many subreddits do not allow posts on unrelated topics, including research advertisement, thus a systematic difference might exist between the types of communities that allow research \textit{v.s.} those that do not. (For example, perhaps norm enforcement is more tightly coupled with SOVC in subreddits that do not allow research.) We intentionally selected diverse subreddits but may have inadvertently missed certain types of communities; future work should periodically revisit our research questions to enrich and update our understanding of the community types we will describe, as well as appending new types we may have missed or which may emerge in the future.

In order to encourage respondents to finish the survey, we limited survey length and did not ask many detailed questions---\textit{e.g.}, about each antecedent individually. We note that respondents provided responses of varying length and detail and that these responses \textit{organically} included information related to antecedents, SOVC, and other community-related concepts. Standard survey limitations apply, including the possibility of misalignment between self-reported data and actual behaviors and opt-in selection bias. Our survey advertisement was posted with different levels of visibility in different subreddits, ranging from  high visibility (\textit{e.g.,} pinned post) to lower visibility (e.g., survey thread). In general, people who elected to take the survey may or may not be representative of the user base of each subreddit. However, respondents from subreddits where there was lower visibility of the recruitment post might be less representative due to finding our study in a more niche way, while respondents who found the survey in a more visible location might be more representative. %Finally, individual users' perspectives as captured in a survey may not fully encapsulate community-wide phenomena; as we will reflect upon in the discussion, future work can confirm or refute the existence of Community Archetypes using group-level techniques.

\subsection{Observation of subreddits}\label{sec:observation}
We visited every subreddit in our sample to observe how users are currently interacting. Rather than programmatically collecting top-voted posts, we scrolled through at least 20 recent posts, along with the comments on those posts, in their order of appearance on the user interface. This technique provides us with a sample of the data near the time when surveys were collected and avoids biasing our understanding of the content toward all-time popular posts (which may not accurately capture the day-to-day diversity of content and interactions on the subreddit). We wrote memos and noted the types of topics users wrote about and whether there were any particular formats or patterns in the content.\footnote{We considered completing a formal, systematic content analysis of posts and comments, however this degree of precision is not required for our research questions, which are better addressed through analysis of survey responses.} We also conducted a preliminary analysis of the types of the rules present in each subreddit to inform our observations; the details of this tangential analysis are included in the supplemental materials.

%For example, we examined whether users were discussing news and whether their posts were formatted in questions. Additionally, we examined the comments under each post to see what kinds of responses were common---\textit{e.g.,} answers to questions, furthering discussions, or little to no responses. 

% \subsubsection{Analysis of subreddit rules}
% In a random sample of 1,000 subreddits, 52\% were found to have any rules at all~\cite{fiesler_reddit_2018}; by contrast, all but one of our subreddits had one or more rules; S7 had none. To analyze all of the rules in these subreddits, the first author coded each rule in our dataset according to the taxonomy of Reddit rule types presented in~\cite{fiesler_reddit_2018}. Every individual rule was classified as \textit{either} prescriptive or restrictive. For all other rule types, codes were applied in one or more categories; most rules were labeled with two or more codes. If a subreddit fit into more than one archetype, its rules were counted in each archetype that applied. Because our sample of subreddits (and therefore rules) is limited, this analysis does not claim a fully representative picture of how rule distributions may appear in each archetype more generally; however, it does provide an impression of how rules vary across our sample. Table~\ref{tab:subredditrules} shows the comprehensive results of this analysis, whereas we highlight a few notable patterns throughout our presentation of RQ2 results (sec.~\ref{sec:users}).

\subsection{Analysis}
We used Grounded Theory Method (GTM)~\cite{muller_curiosity_2014} conducted virtually by the two co-first authors. For the interviews, we used transcripts automatically generated by Zoom as a basis, and then corrected errors while re-listening to recordings. We wrote memos and notes during data collection and transcript correction in order to capture our immediate reflections and ideas about the data. Sensitized by concepts from prior literature, while also allowing new concepts to emerge inductively from the data~\cite{charmaz_constructing_2014}, we then systematically open-coded all interview transcripts. Over the course of approximately two months, the two co-first authors met weekly to analyze and collaboratively cluster all open codes, continuously discussing any uncertainties or disagreements until both authors agreed upon a final clustering arrangement and placement for each individual code. We conducted this affinity mapping using Miro software to first cluster all interview open codes into major axial themes, each with several sub-themes. Similarly, we next open-coded all survey responses and clustered the survey codes in a separate space on the same Miro board. After clustering the interview and survey codes independently, we then iterated upon our clusters and rearranged the Miro board to highlight overlapping ideas between the two groups. Overall, this process generated a total of 1,149 open codes, which we organized into 45 major axial themes with 83 sub-themes. We report on approximately half of these themes/codes that directly address our research questions on SOVC by synthesizing them within the framework presented by this paper.

Importantly, a variety of users' ideas generated distinct clusters that had not emerged prominently in our mapping of researchers' data---thus leading to our conceptualization of these clusters as five ``Community Archetypes'' (see Table~\ref{tab:archetypes}). Upon closer inspection of these clusters and the participant data they were derived from, our discussions inductively revealed that there were clear user groups specific to each archetype for which we assigned the title of ``roles;'' these roles directly led to the interaction patterns which resulted in users' experiences of SOVC. Finally, we cross-referenced the archetype clusters with our subreddit observations in order to assign each subreddit in our sample to one or more archetypes (see Table~\ref{tab:subreddits}). The two co-first authors conducted every step of this analytical work together, and then discussed and edited our presentation of results with all authors for clarity and concision.\footnote{We collected consent to participate in this research according to our exempted IRB protocol, however we did not request permission in advance from participants to make interview transcripts and survey data publicly available. Our researcher data would be impossible to de-identify given the nature of our questions. If future researchers are interested to re-analyze these data, our research team will gladly provide assistance in submitting appropriate IRB protocols for re-use of the data.}

%% file: 4_results_RQ1.tex
\section{RQ1: Researcher conceptions of community \& how to measure it}\label{sec:researchers}

This section seeks to address two central lines of inquiry: (Section~\ref{sec:researchersconcept}) When can researchers understand whether a particular subreddit is experienced as a community or not?;  (Section~\ref{sec:assessing}) What types of behavioral data traces may indicate how and whether users are experiencing SOVC (or not)---either within a particular subreddit (or across Reddit more broadly)?

\subsection{Researcher conceptions of Reddit communities}\label{sec:researchersconcept}
Despite frequently calling subreddits \textit{``communities,''} our participants have not specifically analyzed whether users experience subreddits as communities. We asked whether they draw upon any literature on the theory or praxis of online communities to inform their understanding and methodological approaches; 8 out of 21 participants (1 in computer science, 3 in humanities, and 4 in social science) mentioned at least one paper (\textit{e.g.},~\cite{preece_online_2000}). However, in most cases, researchers have made a certain set of assumptions to frame and constrain their studies, and have used anywhere from one to thousands of subreddits as a source of data for their projects. %Reddit itself refers to each subreddit as a community, making it easy for researchers to refer to them as such without importing external analytic frames. As R6 put it, \textit{``to define a community very simply in Reddit terms, I would say it's a subreddit.''} Aside from \textit{``community,''} 
%Researchers also use many other terms to refer to subreddits. For example, R6 and R11 chose the word \textit{``forum''} because of the post-and-response format of the platform. Researchers with an interest in linguistics used terms like \textit{``discourse community''} (R12) or \textit{``linguistic setting''} (R1). The most popular alternative to \textit{``community''} was \textit{``interest group''} (\textit{e.g.}, R9, R10). 
%Researchers expressed a marked difference between communities and interest groups, with R12 noting, \textit{``an interest group or coalition comes together for a limited set of purposes, whereas a community gathers where at least one of the reasons for gathering is simply to gather.''}
Although some researchers preferred terms like \textit{``forum''} or \textit{``discourse/linguistic setting''}, consensus across participants' responses indicates that it is reasonable for researchers to refer to all subs as \textit{``topical affinity groups''} or \textit{``common interest groups.''} 

Moreover, most researchers made it clear that they do not believe all subreddits are actually communities. As R12 noted, subreddits are simply the \textit{``neutral descriptor of the actual digital space.''} In order to qualify as a community, there must exist some additional set of criteria beyond the content alone, such as: participation in conversations or communication networks, formation of meaningful relationships between users (even occasionally extending into off-platform activities), users' sense of belonging in what they perceive as a community space, and willingness to act on shared values. Many researchers disagreed with a binary approach to whether a subreddit is a community, instead favoring a spectrum model where \textit{``you can have more or less [community] and there's no threshold''} (R21). Others provided further nuance, pointing out that a subreddit could be one big community, but it might also contain \textit{``pocket communities''} (R12)---\textit{i.e.}, smaller communities within larger spaces which do not encompass every user. Although researchers agreed that deciding how to measure or delineate community is quite challenging and imperfect, the next section offers synthesized insights toward considering and assessing SOVC antecedents.

%Other researchers noted the crucial aspect of . Still others noted users' . Ultimately however, researchers seem to agree that \textit{``the people that make it [a community] are the people who participate in it''} (R6). 

% \subsubsection{Summary} To succinctly emphasize the main takeaways of this section, data from our researcher participants suggests that: 
% \begin{enumerate}
%     \item All subreddits can be considered topical affinity groups.
%     \item Individual subs may or may not be experienced as communities, but this is not a binary classification task as much as it is about identifying where a subreddit lies along a spectrum.
%     \item There must be some element of connection beyond the content alone, yet deciding how to measure that is challenging and unclear.
% \end{enumerate}

\constructs

%Because SOVC is internal to and variable across individuals, we cannot directly infer SOVC from observable data. However, in the next section, we summarize researchers' suggested strategies for understanding and measuring four antecedent constructs to SOVC: interactivity, membership, homogeneity, and norm enforcement. %Given that some of these strategies are already utilized in research, we do not claim that they are all novel. Rather, we collate this information here as a concise educational ``mix-and-match'' toolbox for future research. However, in our discussion we describe novel \textit{applications} of these tools, in order to better align research methodologies with users' lived experiences of SOVC. We begin with the construct of \textit{interactivity}--a topic of elevated prominence across our conversations.

\subsection{Researchers' strategies for assessing SOVC antecedents}\label{sec:assessing}
Here, we summarize researchers' suggested strategies for understanding and measuring four antecedent constructs to SOVC: interactivity, membership, homogeneity, and norm enforcement.Although this portion of results is largely recapitulating known methods, it should function as a compact and useful reference point for researchers who would like to design methods in more conscientious alignment with the Community Archetypes we will soon introduce in section~\ref{sec:users}.  

\subsubsection{Interactivity}\label{sec:interactivity}
We identified two major ways that researchers can quantitatively operationalize interactivity, the first of which is more common in the literature despite being less relevant to users' experience of SOVC than the second: 

\begin{enumerate}
    \item \textbf{Volume of observable interaction:} \textit{``A community is nothing without the interaction of the users.''} (R6) In line with this, researchers often measure the level at which users participate---\textit{e.g.}, the aggregate volume, rate, and topic matter of posts, comments, and votes. An important caveat is that the volume of interactivity alone is an overly blunt instrument. A certain level of interaction is necessary, yet insufficient, to indicate SOVC.
    \item \textbf{Degree of conversation:} Researchers agree that there must be \textit{``some robust, sustainable form of communication''} (R16) in order for a subreddit to qualify as a community. People participating in communities are more interactive and discursive than their non-community counterparts; the way to measure that difference is to look for people exchanging back and forth, rather than counting individual comments without interaction between commenters. Researchers should look to identify nested comments and/or repeated user interactions across multiple comments.
\end{enumerate}

\subsubsection{Membership Boundaries}\label{sec:membership}
Deciding how to count users as \textit{members} of a community, versus those who are not, is often a pivotal methods decision with important implications for results. 
\begin{enumerate}
    \item \textbf{Membership tiers:} Participants often assign users to tiers based on: the amount of individual actions taken by the user (from complete inaction (lurking) up to moderating or engaging heavily with the moderation team to advocate for change); the number of posts, comments, and/or amount of time a user spends on the subreddit; or directly asking how users view their position in the community (\textit{e.g.,} in surveys, interviews). %In quantitative studies lacking an opportunity to ask every user, researchers should examine both activity level and time spent on the subreddit to create thresholds that make sense for their project; our discussion will unpack how to reason about these decisions in greater detail, based on what \textit{type} of communities are being studied.
    \item \textbf{Prolonged participation:} Users who return repeatedly to a subreddit understand and develop community culture. These core members help a subreddit maintain its distinctiveness and attract newcomers. Researchers can identify these users by searching for users that post and/or comment a disproportionate amount and have been active on the subreddit for a prolonged amount of time.
    \item \textbf{Linguistic membership boundaries:} Tailored uses of language are also helpful for differentiating members \textit{v.s.} non-members. If users can comfortably use the specific language of a given subreddit, then they have both been there long enough to be familiar with the vernacular, and they are contributing relevant content. Users exhibiting markers such as in-jokes, industry- or topic-specific terms, or particular abbreviations, spellings, or even new words that do not tend to appear elsewhere, are more likely to be members. Researchers also may evaluate whether a user adheres to the style of discourse, including tone, shared opinions, and overall sentiment. %One step beyond adhering to discourse is to take a part in \textit{shaping} it. For example, when a user is an active or important enough member to take the conversation further, for example, by introducing a new in-joke or topic of discussion, then they may be more of a core member.
\end{enumerate}

\subsubsection{Homogeneity}\label{sec:homo}
Measurements of homogeneity will be highly dependent on the researchers' purposes for the project, and the type of subreddit being studied. We discuss four types of homogeneity in ascending levels of perceived utility or accessibility for research.
\begin{enumerate}
    \item \textbf{Demographic:} Unless demographics (\textit{e.g.,} race, age, gender, education, \textit{etc}.) are directly related to the research question, researchers warned against collecting or inferring too much information from demographic homogeneity because there are, likely more often than not, hidden confounds behind demographics than can lead researchers toward faulty claims if they attribute results to demographics rather than the underlying confounds.
    \item \textbf{Situational:} Situational homogeneity considers the user's shared life experiences or situational circumstances in relation to others in the subreddit. Situational homogeneity can sometimes be related to demographics, but it is a more powerful and unifying concept, and is especially important in subreddits focused on events in one's personal life.
    \item \textbf{Goals or values:} Homogeneity in shared goals/values can illustrate why people are in a subreddit and what they hope to accomplish, and is of very high utility for research. It can deviate from other kinds of homogeneity because it is often better to have people with quite different identities and life experiences coming together under a certain goal. %Most researchers talked about observing this type of homogeneity in a qualitative way, such as observing posts and comments over a period of time and determining whether a shared goal or value system exists.
    \item \textbf{Linguistic similarity:} The degree to which a single user's language use adheres to group-level patterns may be useful for indicating their own \textit{individual} membership status. However, assessing the degree to which \textit{groups} of users use similar language may indicate a form of linguistic homogeneity that can be directly calculated from trace data.
\end{enumerate}

Given the absence of demographic information on anonymous accounts, and an inability to accurately infer users' personal situations or goals/values, surveys or interviews should always remain the gold standards for assessing those three types of homogeneity. Qualitative methods with smaller sample sizes can be useful for first deriving the types of personal situations or goals and values that are most relevant to later include in surveys. Ethnographic observation of posts and comments over a period of time can also help to determine what types of situational or shared goals/values exist. Linguistic homogeneity, on the other hand, does not rely upon users' internal states, and may therefore be a convenient metric for studies that cannot directly query users.

\subsubsection{Norm Enforcement}\label{sec:norms}
Norms vary substantially across Reddit, as do the ways they are enforced. Researchers described two broad ways of assessing this.

\begin{enumerate}
    \item \textbf{Vertical:} Vertical norm enforcement refers to top-down moderation by appointed moderators. Researchers suggested quantitative methods for assessing the degree of vertical moderation occurring, such as counting the number of moderators and their volume of activity in the sub\footnote{Note that moderators can also participate more casually, such as by adding their commentary to a discussion, apart from official moderation. A refinement of this strategy requires classifying the purpose of moderators’ messages.}, measuring the proportion of removed content, frequency of banning users, and volume of activity by governance bots, or observing how often rules are modified.
    \item \textbf{Horizontal:} Although users have less power than moderators, they can nonetheless enforce norms horizontally. They can reply to others' posts or comments to point out norm violations (\textit{e.g.,} inappropriate behavior or irrelevant/disallowed content, sometimes even suggesting an alternate subreddit) or report issues to moderators. One strategy for assessing this could be to develop classifiers for identifying this behavior and assessing its frequency. A lack of engagement (\textit{e.g.}, posts receiving no comments) may indicate content that doesn't align with subreddit norms. Finally, up/downvoting might also be an enforcement mechanism, although the meaning of a vote is subjective and not universal, making it an unreliable measure.
\end{enumerate}

Having described researchers' concepts of community on Reddit, and their ideas for evaluating the constructs of interactivity, membership boundaries, homogeneity, and norm enforcement, we will next explore users' perspectives. By juxtaposing researchers' ideas with users' experiences, we can refine and specify assessment strategies for a more human-centered understanding of SOVC.

%% file: 5_results_RQ2.tex
\section{RQ2: User experiences of sense of virtual community}\label{sec:users}

In this section, we present our analysis of our surveys of Reddit users. We find evidence of several broad \textit{qualities of community experience} that are likely to apply across any community. These qualities align closely with prior work and researchers' ideas; therefore we start by briefly highlighting this overlap (sec.~\ref{sec:qualities}). Importantly, we also find that users attribute their SOVC to specific aspects of community experiences that are closely related to specific \textit{types} of subreddits. Researchers' comments largely did not capture these ideas, thus we introduce the framing of \textit{Community Archetypes} and focus most of this section on detailed characterizations of the core ways in which users experience SOVC differently across different archetypes (sec.~\ref{sec:communityarchetypes}).

\subsection{General qualities of community experience}\label{sec:qualities}
There are three major ways users' responses align with approaches suggested by both our researcher participants, as well as by prior work in the social sciences and HCI:
\begin{enumerate}
    \item \textbf{Socializing} is a major factor that improves SOVC; users frequently referenced building meaningful relationships through discussions on threads, iterating on each other's ideas or jokes, and engaging in other subreddit traditions, offline activities, or affiliated online spaces. This user data strongly validates researchers' discussions of the importance of conversation between users rather than volume of interaction alone (sec.~\ref{sec:interactivity}).
    \item Users referenced \textbf{regular participants}, or other users with special flairs as core members. Aligning with prior work and researchers' comments on prolonged membership (sec.~\ref{sec:membership}), it is clear that the visible activities of core community members indicate SOVC.
    \item Users pointed to \textbf{shared experiences, identities, mindsets, values, or views} that bind them together as an in-group. Their comments closely mirror researchers' discussions of homogeneity (sec.~\ref{sec:homo}) and suggest that demographics are not especially relevant, however shared situational circumstances, and shared goals or values, are essential. 
\end{enumerate}
 
We also found that users' perspectives highlighted how the specific configurations of affordances, rules, norms, and behaviors within specific types of subreddits caused them to experience SOVC in unique ways, leading us toward the concept of Community Archetypes.

\subsection{Community Archetypes}\label{sec:communityarchetypes}

\archetypes

Prior works such as~\cite{zhang_working_2022,seering_metaphors_2022} have introduced the language of archetypes to describe conceptual frameworks for how an abstract set of qualities may apply across individual instances. Rather than hard classification categories, boundaries between archetypes are fuzzy, yet each nonetheless retains a unique flavor that may manifest across various identifiable subreddit features and user behaviors. Whereas the authors of~\cite{zhang_working_2022,seering_metaphors_2022} apply the term archetype to individuals, we use it to reference \textit{communities}. This terminology thus helps us to understand how abstract community characteristics lend themselves to certain roles and functions in a user's content consumption or other forms of participation. Given that subreddits are the selected unit of analysis for this paper, we apply these archetypes as labels to individual subreddits, while also suggesting that future work should apply them to other units of analysis on Reddit---\textit{e.g.,} groups of subreddits, or subgroups of users---or moreover, on other platforms that host online communities. 

Our work reveals five community archetypes (Table~\ref{tab:archetypes}). Multiple archetypes may overlap in any subreddit, however, there is typically a more distinct focus on one over others. For example, Table~\ref{tab:subreddits} designates which archetypes best match each subreddit in our study, and shows how some embody one major archetype, whereas others have multiple. Next, we describe each of these archetypes in detail, specifying important \textbf{user roles}, \textbf{content patterns}, and \textbf{rules} associated with each. Table~\ref{tab:archetypes} includes an overview of archetype roles and content patterns.\footnote{Section 4 in the supplemental materials includes more detailed results of our analysis of subreddit rules.}

%\subredditrules

\subsubsection{Topical Q\&A}\label{sec:QA}
Some subreddits are explicitly set up with strict rules to create a Q\&A format around specific topics. For example, P335 declares that S10 is \textit{``the best place on the internet to find reliable, well-sourced, in-depth information on various [academic discipline] topics.''} In our sample of Q\&A subreddits, 55.2\% of rules relate to content/behavior (mainly around respectful engagement with others and following the Q\&A format), with advertising and commercialization (which allows for users to ask for help without being targeted as consumers), low-quality content (which encourages generating content that is useful for the subreddit), and off-topic (which keeps content relevant) rules being the next largest categories, accounting for 10.3\% each. We observe the content pattern that most top-level posts are questions, while most comments are answers, or discussions and enrichments of others' answers. There are two clear user roles: \textbf{novices} (those who ask questions) and \textbf{experts} (those who answer them). These groups may possibly overlap---\textit{i.e.,} a user who usually answers questions might occasionally ask one, or vice versa. Our survey respondents typically indicated their motivations for using the subreddit in an either/or manner; the distinction between these two roles appears to be rather sharp. Expert users enjoy helping people and sharing their knowledge about a subject they are passionate about, while novices sincerely appreciate the special access to experts and the information they can provide. For example, P333 visits S10 to \textit{``provide answers where I have specific knowledge or expertise,''} whereas in S3, P53 enjoys that \textit{``questions can always be answered constructively.''} 

Users explain that SOVC arises when the community is effectively fulfilling its Q\&A function, providing rich, multi-faceted discussions on current topics, while also generating a useful archive of past inquiries and information. For example, P337 appreciates the collaboration and diversity on S10 when \textit{``people with different training and background will often read the same question in a different way and bring different perspectives.''} However, some users, such as P163 mentioned that they don't feel SOVC in S4 because it isn't \textit{``used in a way that could foster a community...it's closer to a help or Q\&A board.''} In general, we posit that each user experiences her own affinity for particular forms of interaction or community; there is no such thing as a one-size-fits-all online community, and it is impossible to claim that SOVC will always arise for every user when $X$ phenomenon occurs. However, across many users, our results suggest that SOVC is more \textit{likely} to arise when $X$ phenomenon occurs, if $X$ is consistent with its associated archetype.
 
\subsubsection{Learning \& Broadening Perspective}\label{sec:learning}
As our most commonly observed archetype, users frequently mentioned learning or broadening their perspective as a primary reason for visiting a subreddit. Content patterns are less formulaic than Q\&A subs, however top-level posts are often pointers to current events, publications, or relevant news, or relatable personal stories, experiences, and questions. Comments tend to elaborate upon the ideas raised, for example by celebrating or disagreeing, making jokes, providing contrasting or similar personal stories and experiences, or adding new ideas and references into the mix. These subreddits have a central focus around a particular culture---not only for people \textit{within} that culture (who benefit from shared experiences of it), but also for users who are \textit{not} part of that culture themselves, yet appreciate it and want to know more. For example, P433 (S11) said, \textit{``I want to hear new and/or different opinions on various topics important to the [redacted] community. It helps me grow my allyship.''} Given this cultural focus, we refer to two roles as \textbf{insiders} and \textbf{outsiders}. For example, on S2, insiders are scientists or people with extensive scientific background and training, whereas outsiders are curious and engaged members of the lay public. On S6, insiders are members of historically marginalized racial groups, whereas other users (mostly white) self-identify as outsiders who want to learn how to be better allies. It may be that this less formulaic, diverse structure is the reason that this category of subs has one of the highest instances of content/behavior rules (62.5\%). Guidance around treating others well and generating content that serves the purpose of the subreddit allows for the content to be more free form while still cultivating a desirable space. They also have higher rates of rules about harassment and hate speech (15\% each) which allows for insiders and outsiders to interface respectfully and in a way that fosters a learning environment.  

Users in these subs report feeling SOVC because of shared mindset or values. For example, P609 (S12) said, \textit{``I think the users of this subreddit are a community because of similar ideas of approach to spirituality.''} Having one big similarity that holds users together allows users to learn from each other's differences. For another example, in S2, users share a value system that places high importance on correct reporting of academic subjects. S2 users have differing expertise and levels of subject familiarity. Therefore, they can inform others who may not have the same background, but who also care deeply about information accuracy. However, some users do not feel SOVC on these types of subreddits because their \textit{``interest in the subreddit is in the content and information more than the interpersonal relationships''} (P341, S10). Such users may not develop SOVC, even if others do, because they are simply not looking for it. 

\subsubsection{Social Support}\label{sec:support}
Much prior work has examined social support behaviors in online communities~\cite{smith_i_2020,andalibi_understanding_2016,andalibi_social_2018}. At a higher level, our work suggests a social support community archetype, in which socially supportive behaviors are the main purpose and organizing principle for gathering. Many users are seeking spaces that can help them navigate their own difficulties or illnesses; others are seeking information and resources to help a loved one who is struggling. For example, P189 said that S7 helps them \textit{``get a better insight as to how to support [their loved one] and what not to do.''} Rules in these subs seek to protect users and keep the conversation focused on the issue, with 61.7\% regulating content/behavior to create a subreddit that is safe and focused, 14.9\% warning against off-topic content (\textit{esp.} content that is unhelpful or triggering), and 10.6\% disallowing hate speech in order to create a safe space. Top-level posts are often specific personal experiences or sensitive disclosures, questions about a health issue, announcements of milestones, reflections or venting, or sharing of resources, artwork, and encouraging thoughts and memes/jokes. These subs often contain flairs that specify content warnings, symptoms, or labels of users' intent (\textit{e.g.,} ``seeking support'' or ``venting''). Comments generally offer support, reflection, commiseration, resources, and validation. In this archetype, roles include \textbf{support seekers} and \textbf{supporters}, yet the distinction is fuzzy and inconsistent, with heavy overlap in how users occupy them. In one post, a user might seek support. In the next, she might become a supporter, returning the kindness she received from others. It is certainly possible to build classification schemes that label individual posts and comments as support seeking or supporting (and sometimes both), however it would be difficult to accurately reflect the role in a static manner per user; it shifts continually. 

Users report SOVC when they see their own hardship reflected in other's posts and the community provides true kindness, support, and validation. P238 (S8) said, \textit{``it feels like my experience was real and valid.''} P189 also said their support community, S5, offers \textit{``sympathy that no one but them can provide properly and meaningfully''} due to other users' intimate familiarity with the issue. Users call these subs a safe space to express feelings, be themselves, and get meaningful help. A few respondents also recognized that they didn't always use these subreddits in a healthy way. P277 often visits S8 for support, but sometimes they  \textit{``want to force myself to get over it by reading about people who have it worse,''} a behavior they do not view as healthy. It is possible to stay too long in support subs, thus promoting lingering in painful memories rather than recovery (unless the user is intentionally returning to help others). It is important to understand when SOVC is beneficial to well-being versus when it could be holding users back.
 
\subsubsection{Content Generation}\label{sec:contentgen}
Users who visit ContentGen subreddits are interested in particular types of content with a certain sense of humor, viewpoint, format, purpose, or type of expression. For example, P4 appreciates that S1 has \textit{``the expectation that you will `act oddly.'''} %Beyond this, users are simply looking for a specific topic and find a subreddit that is devoted to that topic. 
In this archetype, user roles include \textbf{producers} and \textbf{consumers}, where producers create top-level posts that exemplify that sub's specific content style, and consumers are specifically there to view and respond to it. Top level posts might be created by original posters themselves, or they may be contributions from other users or platforms that the poster would like to share or discuss on the subreddit. Comments often include peoples' opinions on the content, extra information about the content, or sometimes commiseration with the original poster. As a result of this content-focused culture, content/behavior is the subject of 62.5\% of rules, low-quality content is 16.7\%, and hate speech and advertisements \& commercialization are 12.5\% each, both of which serve to focus users solely on the content rather than things like ads or hate speech.  

Some users expressed feeling SOVC due to the niche nature of the content which brings people together because, as P271 (S11) states, \textit{``We share the same sense of humor and appreciate the same content.''} Another user, P6 illustrates how niche subreddits like S1 provide a kind of camaraderie that doesn't exist for them in real life: \textit{``It's like finding out there's a `splashing ice-cold water in people's faces' club after years of failing to convince your family and friends that being splashed in the face with ice-cold water is fun.''} However, similarly to the Learning \& Perspective Broadening archetype, users may not feel SOVC if they are simply using the subreddit as a resource rather than looking for community. P142 (S3) illustrated this point, saying, \textit{``It feels much more transactional to me... I do not frequently see updates or multiple posts from a given user.''} 

\subsubsection{Affiliation with an Entity}\label{sec:affiliation}
Some subreddits have explicit affiliations with particular entities, such as geographical places or organizations (cities, universities), popular media (book or fan series, TV shows), sports teams, \textit{etc}. In our limited data set, we only observed this archetype once, in affiliation with a particular geographical area and university campus (S9). Consequently, we observed user roles of \textbf{current}, \textbf{prior}, and \textbf{future} campus residents. However, we will refer to these roles more broadly as \textbf{``affiliates.''} In our usage of Reddit beyond this sample, we have informally observed these affiliate roles across different types of entities; broadly, they can be defined as a user's level of active, ongoing knowledge and investment in a particular entity. Current affiliates have a good grasp on up-to-date ``local'' knowledge by virtue of ongoing attention to or residency within the affiliated entity. Future affiliates are more like ``prospective community members'' who are curious, while prior affiliates enjoy retaining a connection even though they are no longer focusing as much resource or attention. In terms of content patterns, posts often feature local or breaking news and events related to the entity. Future and current affiliates often pose questions, while past (and other current affiliates) tend to answer them. In the comments, users generally express their feelings about news or events, and answer questions or offer advice and opinions about the entity. There is notably less regulation on content/behavior than in other archetypes (40\%), possibly because it is already constrained by the geographical affiliation (however, this may also be because only one subreddit in our sample fits this archetype).

In S9, the specific tie to a geographical entity makes users feel SOVC. P286 (S9) said, \textit{``[City] as a whole has always been a community to me and this is a microcosm of that community.''} Users such as P307 (S9) also appreciate \textit{``people sharing their experiences and advice in a genuine and helpful way,''} because it's likely to be applicable due to users' geographic closeness. Some users felt it was easier to get good information from the subreddit because the city and university administration did a poor job of circulating it. Thus, the subreddit seems to supplement the offline space it is tied to. As P297 put it, \textit{``everything posted there [S9] has to relate to [US university], which helps me feel more connected to the offline campus community.''} In fact, P305 also noticed that \textit{``some memes from [S9] leak onto campus every now and then.''} However, several users stated that this kind of subreddit isn't the place they would go to feel SOVC. P297 (S9) said, \textit{``If I wanted to feel like part of a community, I wouldn't be looking for it on a subreddit for a large state school.''} Another user, P310, remarked that a subreddit like S9 is \textit{``more like a collage of experiences''} than a community.

%% file: 6_discussion.tex
\section{Discussion}
In this paper, we set out to understand researchers' mental models of community, juxtapose these models against users' experiences, and produce recommendations for refining research methods to better align with users' experiences of participating in online communities. Using the Reddit platform as a case study, we found that researchers tend to view subreddits as topical affinity groups, however each subreddit individually sits somewhere along a spectrum of ``community-likeness.'' This result evokes work by \citeauthor{bruckman_should_2022}, who has suggested that groups be categorized as communities on a spectrum based on how closely they resemble a baseline or ``prototypical'' concept of a community~\cite{bruckman_new_2006,bruckman_should_2022}. However, that introduces the problem of what exactly the baseline idea of a community should be. Our work suggests that there is no single prototype that can describe all communities. Rather, our analysis reveals at least five \textit{Community Archetypes} that can be embodied by online communities to varying degrees: Topical Q\&A, Learning \& Broadening Perspective, Social Support, Content Generation, and Affiliation with an Entity (\textit{see} Table~\ref{tab:archetypes}). Each archetype is distinguished by specific behaviors involving user roles, content patterns, and group rules, and users are most likely to experience a sense of virtual community (SOVC) when these behaviors are \textit{strong} and \textit{frequent}. Although we will use Reddit-specific terms and affordances throughout much of our discussion, we will conclude with reflections on how we expect these archetypes to persist across other online social platforms, given the fundamental psychological and social needs shaping human behavior.

\UIcomparison

Our results suggest that subreddits can embody one or two archetypes, however, no single subreddit can fully embody \textit{all} archetypes. To illustrate this, we will compare two popular subreddits: \texttt{r/aww} and \texttt{r/wallstreetbets}.\footnote{We picked large and well-known subreddits in order to illustrate how our framework can impact methods decisions in future studies. We chose these examples because they were \textit{not} included in our sample and they are already well-known among Reddit users, reducing possible exposure of less well-known subreddits to unwanted attention~\cite{smith_disseminating_2020,dym_social_2020}.} Figure~\ref{fig:UIcomparison} depicts screenshots of the subreddits, while Figure~\ref{fig:comparison} contains a hypothetical spiderplot depicting how Community Archetypes may be present within them to differing degrees. \texttt{r/aww} was a ``default sub'' until 2017 when Reddit discontinued this automatic subscription mechanism for new users~\cite{reddit_default_2017} and is described as \textit{``a place for really cute pictures and videos.''} Most rules are restrictive against any other content, and every single post features a picture or video of a cute animal. Thus \texttt{r/aww} appears to be well-described almost exclusively by the Content Generation archetype (possibly with a hint of Learning \& Perspective Broadening). In \texttt{r/wallstreetbets}, on the other hand, we observe a dominant affiliation with finance and stock trading.\footnote{``Wall Street'' is a metonym using the road in Manhattan where the New York Stock Exchange is located to refer more generally to financial markets. \texttt{r/wallstreetbets} was became influential in 2021 due to driving massive upswings in prices of ``meme-stocks.''~\cite{lopatto_how_2021,banerji_reddits_2022}} The community description (\textit{``Like 4chan found a Bloomberg Terminal''}) is more cryptic than \texttt{r/aww}, and the rules focus on a variety of activities that are tolerated (or not) related to stock purchases. Content patterns are strongly evocative of the Affiliation and Learning archetypes, often revolving around discussions of specific stocks and whether users should ``bet'' by buying risky stocks for humor, thrills, or financial gain. There is also an edgy flavor of Social Support in which users who lost large amounts of money on failed bets are simultaneously made fun of, consoled, and glorified. Although a few posts resemble Q\&A or ContentGen, these are not nearly as consistent as in \texttt{r/aww}. These examples highlight how some subreddits have a single or dominant Community Archetype, whereas others have more hybrid, flexible, or nuanced Community Archetypes.

\subsection{Implications for community-centered data science}\label{sec:implications}

We turn our attention first to translating our findings for a data science audience. We will present three reflections on prior work, pointing to implications where Community Archetypes can help to better leverage insights from organizational psychology and more closely align data science with users' SOVC.

\subsubsection{Rethinking ``community size'' as a success metric}\label{sec:communitysize}
HCI papers often use \textit{community size} as a success metric (\textit{e.g.},~\cite{tan_tracing_2018,cunha_are_2019,kairam_life_2012}) on the assumption that the larger a community has grown, the more successful it is. This metric \textit{does} indicate how effectively an online group has recruited users, and the recruitment of a ``critical mass'' of users is a necessary pre-condition and known challenge for community survival~\cite{kraut_building_2012}. However, our work suggests that community size is too blunt of an instrument for assessing SOVC. Rather, users often participate in small subreddits to meet their preferences and needs~\cite{hwang_why_2021} and if subreddits grow too large, studies have noted user complaints that they may have been ``\textit{``better when they were smaller.}''~\cite{{lin_better_2017}} As in Usenet's ``Eternal September''~\cite{grossman_netwars_1997}, influxes of new users may seriously disrupt cherished community dynamics, unless there are sufficient sociotechnical affordances and policies to maintain established practices~\cite{kiene_surviving_2016}. For example, \texttt{r/wallstreetbets} rapidly attracted several million new users after its meme-stocks disrupted the market in 2021~\cite{alcantara_reddits_2021}. By 2022, longtime members felt that \texttt{r/wallstreetbets} lost its distinctive character, since new users' discussions mostly revolved around old, tired topics~\cite{banerji_reddits_2022}. On the other hand, \citeauthor{lin_better_2017} showed that when subreddits were added to default subscription lists by Reddit admins (as was the case in \texttt{r/aww}), there were \textit{momentary} fluctuations in community activity following newcomer influxes, but the subreddits more-or-less returned to pre-established patterns following the disruption~\cite{lin_better_2017}. Newer work similarly explores community resilience after a subreddit is featured on \texttt{r/popular} (Reddit's replacement for default subreddits), finding that community behaviors were differentially and strongly impacted in smaller rather than larger subreddits~\cite{chan_community_2022}. These examples show that community size does not predict SOVC, even though it can \textit{impact} SOVC.

Our work suggests that alternative metrics may be better aligned with users' SOVC. As reported in Section~\ref{sec:qualities}, data scientists should consider community success metrics derived from behavioral traces such as: (1) the frequency and depth of conversations occurring among users; (2) the interactions between well-established community members and newer community members; or (3) the persistence of users' engagement over time. %(Users' shared goals and personal experiences are also promising indicators of SOVC, but these are not inferable from trace data.) 
These strategies should be widely applicable across online communities beyond Reddit as forms of interactivity and membership that are antecedents to SOVC. This concept is supported by prior work demonstrating that greater thread depth on Reddit is associated with language markers indicating positive community outcomes like stability, cohesiveness, and sociability~\cite{mcewan_communication_2016}. However, the Community Archetypes framework also suggests that more specialized forms of interactivity are expected within specific archetypes, and that other antecedents like membership boundaries, norm enforcement, and homogeneity may be quite distinct between archetypes. Our next reflection explores how to operationalize this specificity by re-purposing methods proposed by prior work.

\subsubsection{Rethinking ``SOVC antecedents'' as universal}\label{sec:non-universal}
% In organizational psychology, researchers have adopted a model of SOVC that distinguishes discrete \textbf{antecedents} to SOVC (\textit{e.g.}, interactivity, membership, homogeneity, norm enforcement, etc.) from the cognitive experience of \textbf{SOVC itself}, as well as from \textbf{outcomes} of SOVC (\textit{e.g.}, identification with the community, trust, community cohesion, etc.)~\cite{blanchard_developing_2020}. In this model, the occurrence of SOVC antecedents reliably leads to SOVC, and the experience of SOVC reliably leads to certain outcomes. In the present study, we focused on antecedents to SOVC that may be detectable in trace data as a suggestive proxy for users' SOVC that is not directly observable.\footnote{We chose not to focus our investigation on \textit{outcomes} of SOVC, even though outcomes of SOVC are also valuable indicators of community health. Like SOVC, these outcome constructs tend to assess internal user states that seem less directly inferable from trace data than antecedents.} 

The Community Archetypes framework suggests that users form SOVC as a consequence of community behaviors that differ across archetypes. In other words, different archetypes have specialized antecedents, and this impacts how researchers should interpret different categories of behavior across different communities. For example, \citeauthor{bao_conversations_2021} describe computational measures for a set of eight pro-social behaviors expected in healthy online communities: information sharing, gratitude, esteem enhancement, social support, social cohesion, fundraising and donating, mentoring, and the absence of anti-social behavior~\cite{bao_conversations_2021}. Although these all seem beneficial at face value, they are not necessarily required antecedents of SOVC across all archetypes. For instance, social support is fundamental to the Support archetype, however it may not be expected, needed, or even appropriate in others. If researchers assume that all healthy communities are marked by high levels of social support, they risk inaccurate conclusions that community spaces where social support is low or absent also lack a sense of virtual community. Therefore, for computational metrics related to community features such as pro-social behaviors~\cite{bao_conversations_2021} or types of relationships between users~\cite{choi_ten_2020}, researchers need to choose \textit{which} measures align with users' normative experiences, needs, and values in that particular community and archetype. In \texttt{r/wallstreetbets}, for instance, the edgy style of social support could easily be mis-categorized or perceived as anti-social if the researcher was unfamiliar with the norms of the community. On the other hand, in \texttt{r/aww}, the purpose of the subreddit is to generate cute memes; social support is not the point. Therefore, low levels of social support would not necessarily indicate poor SOVC in this ``ContentGen'' Community Archetype. 

% Moreover, our work raises the the interesting questions: \textit{Are certain archetypes more likely to foster SOVC than others? If so, which specialized antecedents are required for which archetypes?} %and how can the design of sociotechnical algorithms, bots, and UI/UX affordances encourage or foster these antecedents
% For instance, participants' comments suggest that Content Generation subreddits like \texttt{r/aww} may be closer to topical affinity groups, whereas Affiliation or Learning subreddits like \texttt{r/wallstreetbets} may be closer to communities. The present work does not provide sufficient evidence to make firm claims, however this is an interesting line of inquiry for future work. Likewise, future work should also reconsider how to best delineate communities on Reddit, as we will next discuss.

\subsubsection{Rethinking ``community boundaries'' across Reddit}\label{sec:communityboundaries}
As in~\cite{smith_impact_2022}, we acknowledge that a single subreddit is not the best unit of community. Rather, there exist mutualistic, overlapping groups of members across subreddits~\cite{teblunthuis_identifying_2022} with similar topics or genealogical relationships~\cite{tan_tracing_2018} and norms or rules~\cite{chandrasekharan_internets_2018,fiesler_reddit_2018}. News events such as the \#RedditBlackout in 2023 due to the introduction of new API fees~\cite{mehta_reddit_2023} and academic studies have also examined migratory events in which groups of users moved between different spaces on Reddit (or off of the platform entirely) due to moderation conflicts~\cite{newell_user_2016,chandrasekharan_you_2017,davies_multi-scale_2021,trujillo_make_2022}. For example, when Reddit banned the hate-based subreddits \texttt{r/fatpeoplehate} and \texttt{r/CoonTown}, many users left the platform, whereas others moved to new subreddits but mostly did not bring their hate speech with them~\cite{chandrasekharan_you_2017}. %, while \citeauthor{trujillo_make_2022} looked at the impacts of quarantining \textit{v.s.} restricting \textit{v.s.} banning \texttt{r/the\_donald} on users' behaviors~\cite{trujillo_make_2022}. 
These studies highlight how the same communities of users exist beyond the structures of individual subreddits and how these communities may or may not reconstitute themselves elsewhere after major events. 

\textbf{The Community Archetypes framework provides new strategies for delineating communities that may help to describe these phenomena and better match users' experiences of SOVC than individual subreddits.} For example, community detection methods that rely upon network structure~\cite{fortunato_community_2016,olson_navigating_2015,datta_extracting_2019} may benefit from incorporating archetypal community features. One possibility is a ``top-down'' strategy in which researchers could assess and label the archetypal composition of each subreddit in their sample, next analyzing the interaction and relationships specific to each archetype separately. This strategy will allow researchers to look at similar behaviors across multiple different archetypes, but to interpret them more coherently within the context of each archetype. On the other hand, a ``bottom-up'' strategy could de-emphasize subreddits as units and instead look primarily to users' interaction patterns and styles to delineate Reddit communities. For example, researchers could define communities by setting participation thresholds or criteria guided by questions like: Which users of which subreddits are regularly interacting with each other (and/or longer term super users) in nested discussions, regardless of which subreddit the interaction takes place in? What types of \textit{roles}~\cite{buntain_identifying_2014} and \textit{norms}~\cite{chandrasekharan_internets_2018} do these behaviors indicate? Within this community of users interacting across subreddits, what archetype(s) best explain their community's activities? Although this strategy may be computationally taxing, it should enable a more fine-grained analysis with better alignment to users' SOVC.

Finally, it is also important to highlight the temporal boundaries of communities~\cite{barbosa_averaging_2016}. Rather than looking at ``all time'' activities across a subreddit, researchers may better align their methods with users' SOVC if they choose time frames that align with members' actual engagements with the community~\cite{geiger_using_2013}. Users often flow in and out of online spaces, taking their ideas and norms with them. This creates ``cohorts''---\textit{i.e.}, groups of people who entered the community near the same time, often as the result of an event that brought their attention to the topic or to the specific subreddit~\cite{barbosa_averaging_2016}. For example, in Affiliation subreddits (\textit{e.g.}, with a sports team or television show) the community may be healthy and thriving, with a great deal of activity during events related to the entity (\textit{e.g.}, live games or the release of new episodes), even if it is dormant during the ``off season.'' Alternatively, even if activity is quite consistent in a given subreddit, our participants also observed that content may start to feel repetitive when discussions return to the same topics over time. For example, in the Learning archetype, outsiders may have similar questions that regularly percolate to the surface, causing cyclical patterns in the content. Given these types of considerations, we suggest that analyzing cohorts of users who were active during the same period will be more aligned with their SOVC within a given archetypal community.

%In this section, we described how the Community Archetypes framework can refine researchers' selections of community success metrics, SOVC antecedents, and community boundaries. Building on these insights, we will next describe step-by-step recommendations for applying the Community Archetypes framework when designing research methods and community support tools.

\subsection{Applying the Community Archetypes framework in research}\label{sec:CAresearch}
\begin{figure}[t]
    \centering
    \includegraphics[width=\textwidth]{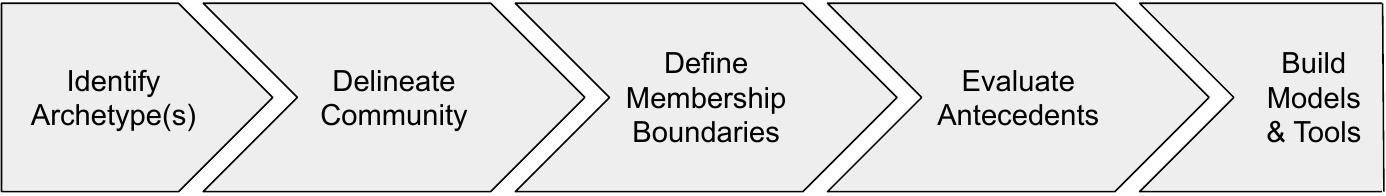}
    \caption{Workflow for applying the Community Archetypes framework.}
    \label{fig:workflow}
\end{figure}

In cases when researchers hope to study, model, or support users' experiences of SOVC in online communities, the Community Archetypes framework can help researchers to structure and justify their selection of methods. This section synthesizes a workflow for doing so (Fig.~\ref{fig:workflow}).

\subsubsection{Identify archetype(s)}
Researchers should first consider which archetype(s) best fit the purposes of their study. (Alternatively, if subreddit(s) are already identified, assessing which archetype(s) those subreddit(s) embody is necessary to analyze them effectively.) Data science models built across many different subreddits without specification of archetype should only rely upon broadly applicable analytical strategies (as in Sec.~\ref{sec:communitysize} above), whereas more sophisticated techniques can be designed if researchers have conscientiously selected particular archetype(s). For methods with smaller sample sizes (\textit{e.g.}, ethnography, interviews, surveys, content analysis, or mixed methods), researchers can use the framing of archetypes, content patterns, and user roles to help structure and scope their protocols, questions posed to users, analytical codebooks, \textit{etc.}, using the archetypal category as justification for asking about certain types of norms, behaviors, \textit{etc.,} and not others.

% On the other hand, if a researcher is looking to build  (\textit{i.e.}, those outlined in \ref{sec:qualities})  that will reliably predict the presence of SOVC. Any other assumptions made my be inaccurate and could give incorrect results. Perhaps the best course of action is to identify sets of subreddits that fit a particular archetype. This way, it is possible to build methods based on more specific criteria and there will be a higher likelihood that assumptions can hold across them. 

\subsubsection{Delineate community}
Once archetype(s) have been chosen, researchers can delineate community boundaries at the intersection of those archetype(s), the topic of interest, and a sensible timeframe. This can be done through examination of groups of candidate subreddits \textit{or} groups of interacting users across subreddits (sec.~\ref{sec:communityboundaries}). Researchers can choose qualitative or quantitative techniques for analyzing content patterns, user roles, and rules/norms (\textit{see} Table~\ref{tab:archetypes}) depending on the size and scope of their study. However, in order to ensure that assumptions hold across their sample, they should select communities of similar archetypal composition. 

Most researchers in our study opted to study the largest or most popular possible subreddit(s) on their topic of interest. However, large subreddits have distinct moderation and governance challenges relative to smaller subreddits. The question of what size subreddit is ``most representative'' or ``more of a community'' is difficult given that long-tailed distributions have no central tendency: are many small subreddits more representative of a user's typical experience or a few large subreddits? The largest subreddit may not always be the best choice, particularly when it is important that a community exists or a specific kind of data is needed. For example, one researcher described how many fandoms have multiple Affiliate subreddits, the largest of which is often for more general information, casual fans, and people just discovering the fandom (\textit{i.e.}, future affiliates), whereas smaller ones are dedicated to more specific sub-topics for much more highly invested current affiliates or have more stringent moderation rules. More community-like behaviors may exist in these smaller, more specific subreddits than in the larger ``on-boarding'' subreddit. Therefore looking at smaller, more specialized subreddits or groups of users interacting across the whole family tree of fandom Affiliate subreddits may be valuable tactics. 

\subsubsection{Define membership boundaries}
The next step is to identify how users should be classified as members. Our framework suggests a tiered membership classification scheme because a binary view of membership may be overly simplistic and provides an incomplete interpretation of actual community dynamics. An ideal way to determine membership in qualitative and mixed methods is to first observe what user roles exist, and then ask users directly about how they view their own membership and role in the community. In quantitative and data science-based methods, rather than a simple activity threshold (\textit{e.g.,} ``members'' are defined by making $X$ number of posts), researchers should identify markers of expected user roles within the specified archetype(s). Prior work on identifying social roles in online communities~\cite{buntain_identifying_2014,zhang_working_2022,muller-birn_peer-production_2015,jin_beyond_2007,zhu_identifying_2011} and techniques that combine behavioral and content features~\cite{liu_identifying_2019} will be especially useful in this regard.

\subsubsection{Evaluate remaining antecedents}
The first three steps set the stage for more effectively assessing the remaining antecedents (interactivity, homogeneity, and norm enforcement) with respect to the appropriate archetype. 

\textbf{Interactivity.} 
%Interactivity is the cornerstone of any community. 
Different \textit{forms} of interactivity are more essential in different archetypes (sec.~\ref{sec:non-universal}). Prior work provides examples of ways to assess some antecedents (\textit{e.g.},~\cite{bao_conversations_2021}). New methods can be tailored to distinctive archetypal content patterns:
\begin{itemize}
    \item \textbf{Q\&A:} Researchers can assess whether almost all posts are formatted as questions, whether the majority are receiving answers, and how in-depth and varied the answers are.
    \item \textbf{Support:} Support exchange is fundamental and SOVC is likely when users regularly switch between support-seeking and support-providing roles.
    \item \textbf{Affiliation:} Frequent references to entity-specific terms, people, and topics encourage SOVC; the community should become more active during important entity-affiliated events.
    \item \textbf{Learning:} SOVC is likely when in-depth, respectful discussions feature many contrasting perspectives between insiders and outsiders.
    \item \textbf{ContentGen:} Posts and in-jokes should adhere to expected content formats and not be repetitive re-posts. Posts that refer to and build on prior content should indicate SOVC.  %Unique and recurring patterns in post content and comments could be identified and the evolution traced back to their origins.
\end{itemize}
 
%These examples are not comprehensive, nor do they do imply that other types of interaction are without value. However, users heavily emphasized these types of interactions when discussing what made their subreddits feel like communities or what could make their communities better. %For this reason, we suggest paying special attention to these interaction types when determining whether community behavior exists. 

\textbf{Homogeneity.} 
Participants suggested that demographic homogeneity (race, sex, age, \textit{etc}.) is less salient than other forms of homogeneity, such as shared interests, experiences, and goals. Even these latter forms may be differentially important across archetypes. In qualitative studies, researchers may want to focus on shared interests for Q\&A and ContentGen communities, shared experiences for Support and Affiliation communities, and shared goals in Learning communities. In quantitative studies, linguistic homogeneity is important to community cohesion and should be useful across all archetypes. %Analyzing linguistic homogeneity can involve: the proportion of positive \textit{v.s.} negative language used; specific dialects; certain words or phrases that are unique to the community; certain tones (supportive \textit{v.s.} combative, informative \textit{v.s.} casual), and others. %Researchers may want to use the archetype to help inform them about the linguistic homogeneity they should be looking for. Importantly, we suggest that researchers not use demographic homogeneity as a classifier (race, sex, \textit{etc}.) unless directly called for by the research question. Otherwise, those identifiers are rarely the root of the cohesion and may miss an entirely different unifier. Even in situations where a person's demographic identity may be a part of why they feel community in a space, it is more about the experiences they have had that are tied to that identity, rather than the identity itself. In this case, the experience is much more important than the identity. [cites]

\textbf{Norm Enforcement.} 
There are \textit{macro} norms (universal to much of Reddit), \textit{meso} norms across particular groups of subreddits, and highly specialized \textit{micro} norms within individual subreddits~\cite{chandrasekharan_internets_2018}. Community Archetypes offers a new lens for interpreting meso norms: we expect that subreddits of similar archetypal composition are likely to be marked by similar meso norms, and that users' SOVC is related to how these norms are vertically and/or horizontally enforced. %Our cursory analysis of the rules in our sample of subreddits (\textit{see} Table~\ref{tab:subredditrules}) is too limited to support broad claims about which exact rules and norms are expected in which archetypes, however 
Future work should explore this question in more depth, particularly by using the taxonomy of Reddit rules by~\citeauthor{fiesler_reddit_2018} (as in our cursory rules analysis included in the supplemental materials)~\cite{fiesler_reddit_2018}. We expect differences across archetypes in terms of: what types of rules/norms exist; how strictly users \textit{want} norms to be enforced; and how strictly moderators enforce norms in actual practice. For instance, users of Support subreddits especially emphasized how their SOVC is contingent upon stricter enforcement and mods' commitment to keeping the space safe; these subs may need more rules overall and higher enforcement both vertically and horizontally to protect users' SOVC. Q\&A and ContentGen subreddits may need stronger adherence to prescriptive content rules, whereas Learning and Affiliation subreddits may need better mechanisms to govern behaviors like harassment, hate speech, and influence operations. %We note that extreme over or under-enforcement of rules and norms can severely undermine SOVC, directly leading fracturing or forking communities who then compete for users and antagonize each other (\textit{e.g.}, brigading). Researchers should carefully consider the methodological techniques in Section~\ref{sec:norms} to explore these issues.

\subsubsection{Build models and community support tools}
Community Archetypes can sensitize our research methods, data models and community support tools to meaningfully nourish the human need for sense of community in our modern world. For example, users told us that strategies like increasing the activity of moderators, keeping subreddits small but fostering more activity and content, and structuring regular activities both on Reddit (\textit{e.g.}, routine threads for catching up with other users) and off Reddit (\textit{e.g.}, on Discord) would contribute to improving SOVC. They also highlighted opportunities for better governance bots~\cite{smith_impact_2022} for patrolling hate speech and troll activities, specific tasks in that subreddit's interest area, answering repetitive posts on common questions, or responding to posts that never received any comments.

\paragraph{Tailoring governance support to Community Archetypes} Similarly to how rules and norms vary by archetype, our work suggests that there is no ``one-size-fits-all'' set of moderation strategies, policies, or tools that are likely to work equally as well for all communities. Styles and types of moderation labor should differ in reliable ways across different archetypes; thus, a useful future contribution  could be baseline tooling tailored appropriately to each archetype. Rather than one \texttt{u/AutoModerator} to rule them all, subreddit mods could be provided with options configured and tuned to more sophisticated patterns of activity in the subreddit---not only for enforcing rules restrictively, but also for spurring positive antecedent behaviors, interactions, and communication that lead to SOVC in that archetype. For example, imagine an \texttt{u/AutoSupportMod} for Social Support subs, an \texttt{u/AutoMemeMod} for ContentGen, \textit{etc.}

\subsection{Community Archetypes beyond Reddit}\label{sec:futurework}

\subsubsection{Archetypes across platforms} 
Reddit is now a prominent platform studied by academics, however that could easily change given the ever-shifting landscape of social media. The archetypes uncovered in our study relate to basic human needs for information, entertainment, socialization, and support, and that they likely feel familiar to readers. Entire platforms with more constrained affordances and contexts of use than Reddit exist specifically for Q\&A (\textit{e.g.}, StackOverflow, Quora), Social Support (\textit{e.g.}, CaringBridge, PatientsLikeMe), Content Generation (\textit{e.g.}, Imgur, MemeGenerator), etc. At the same time, platforms like Facebook, Instagram, Twitter, or analogous Fediverse options can support similar Community Archetypes depending on how users and/or moderators appropriate them. Reddit communities likely learned a thing or two from their predecessors, just as future platforms will iterate on today's models. We expect this Community Archetypes framing to remain fundamentally relevant to research because the underlying community needs will remain relevant to users. Yet future work will need to account for new sociotechnical circumstances, such as issues raised during the \#RedditBlackout in June 2023.

Although Reddit's API was free and open for many years, new API rules have restricted access to data, impacting researchers, developers, and users. Moderators of thousands of popular subreddits made their subs private (and therefore inaccessible to millions of members) or participated in other forms of protest such as bizarre or humorous content restrictions~\cite{mehta_reddit_2023}. Although Reddit granted mods a limited tier of free API usage for running moderation bots~\cite{reddit_help_moderation_2023}, it also threatened to remove mods who refused to re-open their subs~\cite{peters_reddit_2023}. Issues like this may cause users to abandon the platform, migrate to other subreddits, or alter their community participation behaviors. Amdist the instability of shifting platform incentives and migrations, Community Archetypes offers a promising way to conceptualize, study, and support real human communities---regardless of the platform they currently inhabit.

%Community Archetypes provides a lens for understanding systematic shifts in participation. For example, some Support subreddits announced that they will not participate in the \#RedditBlackout because they recognize that users truly need the support (\textit{e.g.,} \href{https://www.reddit.com/r/CPTSD/comments/147dyz9/announcement_rcptsd_not_joining_the_blackout/}{r/CPTSD}~\cite{rovinrockhound_announcement_2023}).}

\subsubsection{Future work} Given the rapid rate at which algorithms, affordances, behaviors, and governance of tech platforms continue to evolve, the Community Archetypes framework is certainly not complete. We did not specifically set out to find these archetypes, thus different archetypes could exist or emerge on other current or future social media platforms. Future work may identify additional archetypes missed by this study, as well as methodological considerations that could differ accordingly. Additionally, there is a need for case studies in which we can empirically use and validate the Community Archetypes framework not only on the level of individual ``digital containers'' of communities (\textit{e.g.,} subreddits) but even more especially across \textit{groups} of such containers that provide users with SOVC. Moreover, trace data can easily be misinterpreted and can never tell the whole story of users' internal states or feelings. In order to truly understand how users are feeling, it will always be necessary to ask them directly, thus future work should continue engaging directly with community members to understand their experiences and perceptions as platforms change.

\subsection{Conclusion}
In this study, we interviewed 21 Reddit researchers and surveyed users of 12 subreddits to explore SOVC. We identified five Community Archetypes with distinctive forms of activity, user roles, and  content patterns. Accordingly, we contribute methodological recommendations for identifying and studying each of these archetypes. This work will support researchers to understand, select, and justify methods that best suit their projects, as well as to design community support tools that can promote healthy community formation and provide members with nourishing SOVC.

%% file: interviewprotocol.tex
\section{Interview Questions}\label{sec:interviewquestions}
We used the following interview questions as the basis for semi-structured interviews, asking additional follow-up questions as required:

\begin{enumerate}
\item How have you personally used Reddit in the past? \item What led you to begin studying Reddit?
\item How would you describe your mental model of what a community is on Reddit?
\item Do you use a specific definition of “community” in your research?
\item What experience or literature (if any) are you drawing from in your conceptualization or definition of what a community is?
\item How do you operationalize this concept of community in your research?
\item In your work, which users do you consider to be (or include as) members of this community? (Or: How do you think about the boundaries of community membership)
\item (If quant) When you’re choosing how to structure predictive or descriptive tasks, how do you select which variables to include?
\item (If quant) How do you assign values to parameters of your operationalization that don’t necessarily have pre-existing “ground truths”? (Possibly: How do you justify your selections of methods or models?)
\item What do you think are the strengths/limitations of your operationalization?
\item Are there any variables you wish you had that you think would strengthen or validate your decisions?
\item To what degree do you feel that you explain these considerations in your papers? Is there anything you don’t write about--e.g., due to space, disciplinary norms, presentation or ethical concerns? 
\item If a subreddit were to have a strong ``sense of community,'' what types of observable characteristics might you expect in that sub?
\item Do you have a sense of what types of features, norms, or mechanisms cause these characteristics?
\item  How would you assess whether different subreddits or communities on Reddit are related or similar to each other?
\item Do you have ideas about mechanisms or reasons that cause users to: form new subreddits? migrate to other subreddits? leave the platform?
\item Anything else you’d like to discuss that I haven’t asked about?
\end{enumerate}

% \subsection{Talk-aloud Questions}\label{sec:talkaloud}
% Each bullet below contains text displayed on one slide to participants:

% \begin{itemize}
%     \item \textbf{Homogeneity:} 
% The degree of similarity of users related to their values, attitudes, goals, or other personal characteristics (e.g. demographics, personality traits, etc.).
% \item \textbf{Membership Boundaries: }
% The degree to which users can be considered members of a community, vs. not being considered members of a community.
% \item \textbf{Interactivity: }
% The degree to which users are interacting within the community.
% \item \textbf{Norm Enforcement:}
% The degree to which norms are enforced within the community.
% \item Are there any other types of features, variables, or behaviors that you feel might be important to users' sense of community that we haven't talked about yet?
% Or anything else you'd like to discuss that you feel is relevant to this study?
% \end{itemize}